# A PRECISION BENCHMARK SUITE FOR NUCLEAR REACTOR POINT KINETICS EQUATIONS VIA CONVERGED ACCELERATED TAYLOR SERIES (CATS)

**B.D. Ganapol**
Department of Aerospace and Mechanical Engineering
University of Arizona
Ganapol@cowboy.ame.arizona.edu

## ABSTRACT

Extreme benchmarks of ten or more places for the point kinetics equations for time dependent nuclear reactor power transients are rare. Therefore, to establish an extreme benchmark, we will employ a Taylor series with continuous analytical continuation (CAC) to solve the ordinary differential equations of point kinetics including feedback. Non-linear Wynn-epsilon convergence acceleration confirms the highly precise solutions for neutron and precursor densities. Through adaptive partitioning of time intervals, the proposed Converged Accelerated Taylor Series, or CATS algorithm [1] in double precision, automatically performs successive mesh refinement to obtain high precision initial conditions for each sub-interval, with the intent to reduce propagation error. Confirmation of ten to twelve places comes from comparison to the BEFD (Backward Euler Finite Difference) algorithm [2] in quadruple precision also developed by the author. We report benchmark results for common cases found in the literature including step, ramp, zigzag and sinusoidal prescribed reactivity insertions and insertions with non-linear adiabatic Doppler feedback. We also establish a suite of new prescribed reactivity insertions and insertions with feedback, based on reactivities with Taylor series as suggested by the CATS algorithm.

## 1. MOTAVATION

The numerical solution of the point kinetics equations (PKEs) characterizing nuclear reactor transients is one of the most noted computations in all of reactor physics. The PKEs admit few analytical solutions, so numerical means are necessary and consequently have drawn attention from nuclear engineers, applied mathematicians and physicists. It is not surprising to find that nearly every well-known method of solving a system of ODEs applied to the PKEs. One only need survey the list of references found in this work and the references therein to verify the truth of this statement. Among the many proposed solutions have been Runga- Kutta (RK) methods of all orders including generalized RK [3,4] and the Rosenbrock forms [5]; exponential transform or basis functions [6]; stiffness integrating factors [7]; Padé approximants [8,9,10]; Laplace transform inversion [11]; piecewise constant reactivity approximations [12,13]; finite difference and converged accelerated finite differences [14,15,16] and application of integral equation [17,18,19].

Such an important set of equations therefore demands adequate numerical standards of solution, or high precision (often-called extreme) benchmarks, to ensure confidence in the numerical algorithms that have been, or will be, proposed. A high precision PKE benchmark, based on Taylor series called Converged Accelerated Taylor series (CATS) [1], appeared in the 2012 American Nuclear Society Mathematics and Computation Topical in Knoxville. Though not in the mainstream, the CATS benchmark managed to contribute to the development of a considerable number of published algorithms since its first appearance. Along with a second benchmark based on BEFD (Backward Euler Finite Difference) [2], the two algorithms are arguably the most precise found in today's literature. My objective therefore, is to bring an updated and more precise CATS version (at least to ten-places) to a wider audience and simultaneously to augment the existing examples of reactivity insertions with new ones.

Our focus is to compute ten-place solutions to the PKEs from a fast, efficient and straightforward Taylor series expansion. PKEs find application in nuclear facilities worldwide as numerical simulators, where they are essential to the design and operation of next generation reactors and research facilities. Aside from their practicality, there is a particular satisfaction associated with developing a reliable, efficient, straightforward and competitive numerical solution for PKEs for a wide variety of transients. "Efficient and straightforward" are to be interpreted as simple, yet extraordinarily precise within a reasonable execution time.

Taylor series solutions to the PKEs [20,21,22,23,24,25] have been prominent and are arguably the most straightforward of all. Notably, the Taylor series was one of the earliest solutions conceived of, but the approach never gained popularity as it was thought to have poor convergence. In addition, time step control is required and the algorithm exhibits rather delicate numerical behavior because of its susceptibility to round off error. In his classical work, John Vigil [23] presented one of the most precise implementations of the Taylor series solution through continuous analytical continuation (CAC). His work was truly ahead of its time. Vigil chose time steps to systematically reduce propagation and round off errors. The CAC concept still requires a time step, since the Taylor series representation is over an interval. However, with reduced time steps, fewer terms in the series are necessary for convergence than otherwise, thus leading to a higher precision evaluation in potentially less computational time. At long times however, the algorithm still suffered from propagation error accumulation, as does any solution of an initial value problem, unless one takes precautions to reduce error propagation, as is done here.

Before addressing the numerical solution, it is natural to question the significance of a high precision benchmark. It seems that five or six places would suffice, which indeed would be most practical. However, my aim is to go beyond practical and to push the boundaries of what is possible. Specifically, to address the question -- Can a Taylor Series (TS) serve as an extreme benchmark solution (nine/ten-place or better) for the PKEs? The CATS algorithm implements the common TS approximation at the numerical margins of double precision (DP) arithmetic rather than with extended precision arithmetic to maintain computational simplicity and to be accessible. The term "extreme benchmark" means to find solutions to nine or more digits of precision by application of readily accessible DP arithmetic. We will explore the essential elements required to address extreme benchmarking including continuous analytical continuation, partitioning, adaptivity, convergence acceleration, numerical "wrangling" and benchmark confirmation.

We begin with the mathematical theory of the Taylor series solution followed by the numerical formulation for high precision. Confirmation of the CATS algorithm then follows for a step insertion demonstrating 23 places of precision. The usual prescribed reactivity insertions found in the literature come next followed by non-linear insertions including Doppler. The last two sections present a suite of new reactivity insertions including both nonlinear prescribed insertions and variations of non-linear Doppler insertions.

## 2. MATHEMATICAL THEORY

We begin with the PKEs for an arbitrary number ($m$) of delayed neutron groups without an external source for simplicity:

$$\frac{dN(t)}{dt} = \left[\frac{\rho(t,N)-\beta}{\Lambda}\right] N(t) + \sum_{i=1}^{m} \lambda_i C_i(t)$$

$$\frac{dC_i(t)}{dt} = \frac{\beta_i}{\Lambda} N(t) - \lambda_i C_i(t), \quad i=1,...,m,$$

(1a,b)

where $N$ and $C_i$ are the neutron and precursor densities and $\rho$ is the reactivity, which may depend upon the neutron density and time $t$. $\beta_i$ is the $i^{\text{th}}$ delayed group yield directly from fission and $\beta$ is the total yield from all groups. $\lambda_i$ is the decay constant for the $i^{\text{th}}$ delayed group and $\Lambda$ is the neutron lifetime. All kinetic parameters except possibly reactivity are constant in time.

Next, we assume a time grid as shown in Figure 1 for arbitrary time edits $\{t_0, t_1, ..., t_j, ..., t_J\}$, except $t_0 \equiv 0$. Neutron and precursor densities at the time edits are the primary numerical objective. If expanded in a TS about the initial time $t_{j-1}$ of the edit interval $t_{j-1} \le t \le t_j$, the neutron, precursor densities and reactivity, assumed to be piecewise infinitely differentiable within an interval, are

$$\left\{ \begin{array}{c} N_j(t) \\ C_{i,j}(t) \\ \rho_j(t, N_j) - \beta \end{array} \right\} = \sum_{n=0}^{\infty} \left\{ \begin{array}{c} N_{n,j-1} \\ C_{i,n,j-1} \\ \rho_{n,j-1} - \beta \delta_{n0} \end{array} \right\} \left(t - t_{j-1}\right)^n. \qquad (2\text{a,b,c})$$

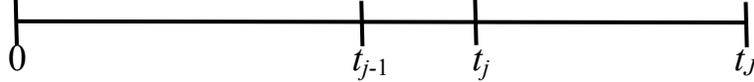

Fig. 1. Time edit grid.

The notation for all TS coefficients is

$$f_{n,j} \equiv \frac{1}{n!} \left. \frac{d^n f(t)}{dt^n} \right|_{t_j} \qquad (2\text{d})$$

including a subscript $j$ to indicate the time interval.

On multiplying Taylor series as a Cauchy product [26], the first term on the RHS of Eq(1a) becomes

$$\left[ \frac{\rho_j(t, N_j) - \beta}{\Lambda} \right] N_j(t) = \frac{1}{\Lambda} \sum_{k=0}^{\infty} \gamma_{k,j-1} \left(t - t_{j-1}\right)^k, \qquad (2\text{d})$$

where

$$\gamma_{k,j} \equiv \sum_{l=0}^{k} \left( \rho_{k-l,j} - \beta \delta_{k-l,0} \right) N_{l,j} = \left( \rho_{0,j} - \beta \right) N_{k,j} + \sum_{l=0}^{k-1} \rho_{k-l,j} N_{l,j}. \qquad (2\text{e})$$

Substitution into the PKEs, with the RHS derivative appropriately represented as

$$\frac{d}{dt}\begin{Bmatrix} N_j(t) \\ C_{i,j}(t) \end{Bmatrix} = \sum_{n=0}^{\infty} n \begin{Bmatrix} N_{n,j-1} \\ C_{i,n,j-1} \end{Bmatrix} (t-t_{j-1})^{n-1}, \qquad (3)$$

gives the recurrences for the TS coefficients for $n = 0,1,...$

$$(n+1)N_{n+1,j} = \frac{1}{\Lambda}\gamma_{n,j} + \sum_{i=0}^{m}\lambda_i C_{i,n,j}$$

$$(n+1)C_{i,n+1,j} = \frac{\beta_i}{\Lambda}N_{n,j} - \lambda_i C_{i,n,j}. \qquad (4a,b)$$

Starting from critical at $t_0$, the starting terms for the recurrence come from the initial conditions

$$N_{0,0} = N_0(0)$$

$$C_{i,0,0} = \frac{\beta_i}{\lambda_i \Lambda}N_{0,0}. \qquad (4c,d)$$

From continuity of the initial edit $t_{j-1}$ of edit interval $j$, the zeroth term of the Taylor series is the neutron or precursor density at the end of the previous time interval $t_{j-1}$

$$N_{j-1}(t_{j-1}) = N_j(t_{j-1}) = \sum_{n=0}^{\infty} N_{n,j-1}(t_{j-1}-t_{j-1})^n = N_{0,j-1} \qquad (5a)$$

$$C_{i,j-1}(t_{j-1}) = C_{i,j}(t_{j-1}) = \sum_{n=0}^{\infty} C_{i,n,j-1}(t_{j-1}-t_{j-1})^n = C_{i,0,j-1}. \qquad (5b)$$

which is the definition of Continuous Analytic Continuation (CAC). A key feature of the CATS algorithm is once one assumes the Taylor series for reactivity either from its functional form or, say from a Padé approximation, the recurrence is complete with all coefficients deriving form Eqs(4a,b). Equations (2)-(4), therefore define continuous analytical continuation (CAC) (connecting edit sub-intervals) to provide the basis for extraordinary precision.

Theoretically, Eqs(2a,b) are the true analytical solution representations-- though not in closed form. Most importantly with Doppler feedback, when the reactivity is expressed as a piecewise continuous TS of the neutron density, the recurrence for the TS coefficients for the neutron density [Eqs(4b,c)] naturally lags the RHS relative to the LHS. Thus, the RHS is always numerically available without approximation and the TS solution in the limit of an infinite number of terms becomes exact. Note the previous statement is true only for piecewise continuous reactivity, which is always satisfied within an interval.

From a purely numerical consideration, there are two numerical errors associated with the TS solution of the PKEs, besides round off error. The first error is the evaluation of the TS itself, or the series truncation error, and is the local error. The TS evaluation requires convergence of its partial sums, which in the limit admit convergence acceleration. The second error comes from the propagation of the local error from one time interval to the next like in a finite difference approximation. From Eqs(5), we observed the solution propagates through CAC and is continuous at an interval end-point and the adjacent interval initial point. Minimization of the local and propagation errors are the essence of the CATS implementation through grid refinement and acceleration as discussed and demonstrated next.

## 3. NUMERICAL IMPLEMENTATION
### 3.1 Continuous Analytical Continuation with Partitioning
As already mentioned, to achieve extraordinary accuracy, we base the TS solution for the neutron and precursor densities on CAC with grid refinement convergence. Note however, we are primarily interested in the neutron density. To do so, one evaluates the recurrence of Eqs(4) on the arbitrary discrete set of time edits $t_j$ forming the edit-or sub-intervals $[t_{j-1},t_j]$ of the (total) interval $[0,t_J]$ as shown in Figure 1. Since the first coefficient initiates the recurrence, all coefficients follow; therefore, enabling construction of the TS of Eqs(2) to converge, in principle, to any precision. The concept of CAC therefore continues the TS solution between sub-intervals by continuity through the last sub-interval. Generally, the smaller the time step (sub-interval) the more quickly the TS convergences since fewer terms are necessary to converge the partial sums. Thus, high precision convergence in a realistic computational time requires time step management. Therefore, we introduce additional partitioned sub-intervals within edit intervals along with convergence acceleration of the approximate solution at each edit $t_j$ (as explained below).

### 3.2 Operational Flow Chart Explained
The CATS algorithm is organized around edit sub-intervals specified at input as shown in the Flow Chart of Figure 2 and further clarified. The desired edit is the end-point of the interval $[t_{j-1},t_j]$, which is determined in Loop **(A)**. However, in general, a more refined interval is required for efficient TS convergence, thus the sub-intervals are refined in Loop **(B)** through further subdivision by partition.

Fig. 2. CATS Algorithm Operational Flow Chart.

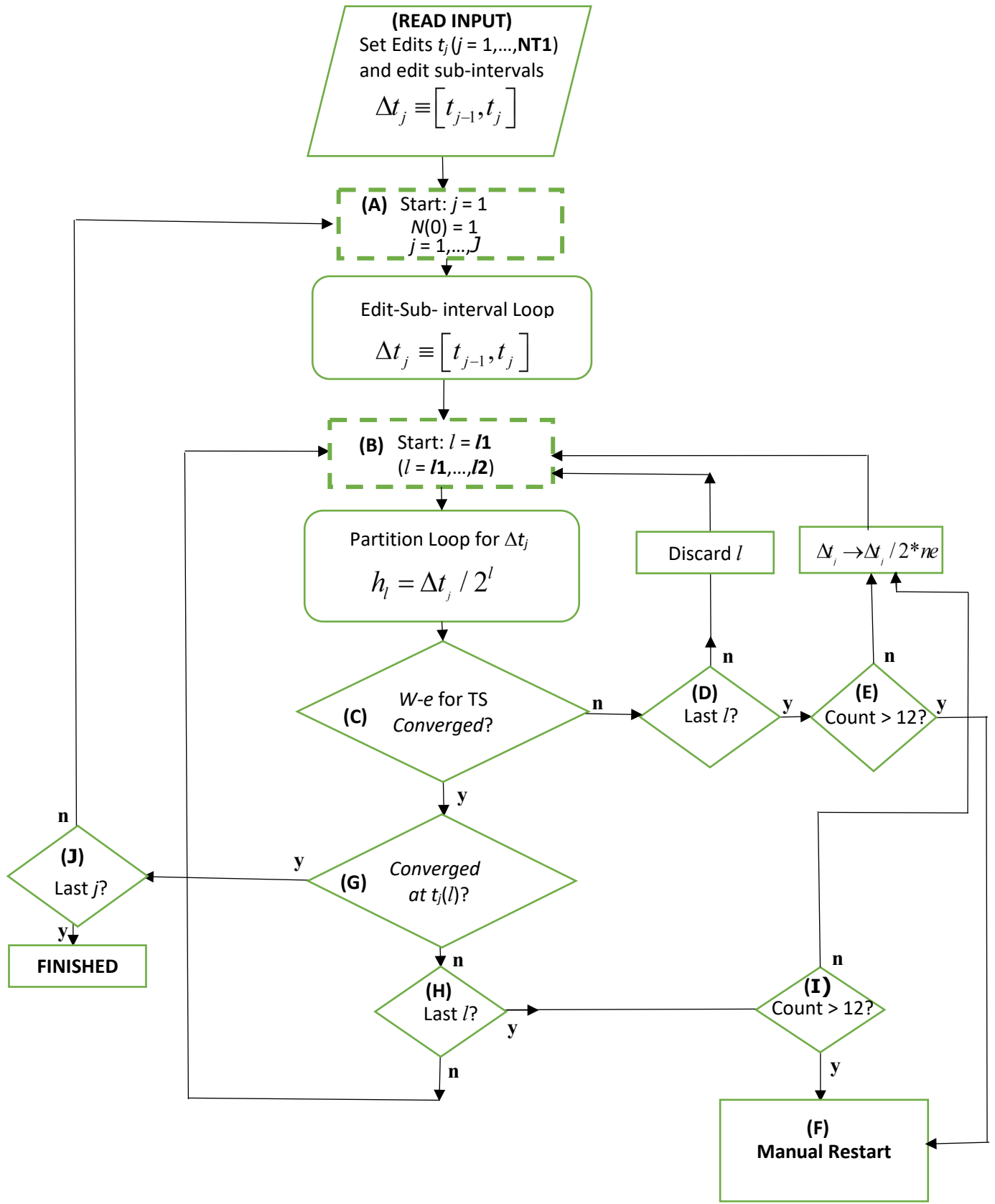

Refinement enables the TS **(C)** to converge more efficiently, either through acceleration or sequentially. Each refinement gives an approximate solution at the edit sub-interval end-point $t_j$ and therefore together form a sequence of solutions for the densities. One monitors the sequence of densities to be either accelerated for convergence by the Wynn-epsilon (W-e) acceleration [27] or allowed to converge sequentially **(G)**.

We now briefly outline the operational flow as shown in the flow chart of Figure 2a

The letters **(XX)** in the diamond boxes refer to decisions with outcome **y** or **n** as indicated. The dashed outlined rectangles indicate the beginning of loops with their indices specified.

Referring to Figure 2, the explanation of the algorithmic flow is as follows:

**(A)** *Edit sub-interval loop*

Loop **(A)** specifies the $j^{th}$ edit sub-interval $\Delta t_j = [t_{j-1}, t_j]$ to be partitioned eventually covering all $J$ intervals over the total time interval $[0, t_J]$.

**(B)** *Partition of sub-interval loop*

The $j^{th}$ edit interval is then sub-partitioned in Loop **(B)** according to

$$^1 h_l = \frac{\Delta t_j}{2^l}. \tag{6}$$

for (partition) $l$ in $[l1, l2]^2$ to determine the densities at edit $t_j$ for partition $l$.

**(A/B/Cn/Dn/B)** *Convergence$^3$ of TS?/Last l?*

If the TS does not converge **(Cn)** at partition $l$ within $[l1, l2]$ of Loop **(B)** in **(C)**, the current we discard partition $l$, increment $l$ and Loop **(B)** partitioning continues unless $l$ is the last $l$ (**l2**), which is not true yet **(Dn)**.

**(A/B/Cn/Dy/En/B)** *Convergence of TS?/Last l?/Count < 12?*

If at the last $l$ **(Dy)**, the partition loop fails to give a converged TS for all densities at $t_j$, the original edit interval is partitioned by 2***ne** and the partition Loop **(B)** restarts for the newly formed partitions as long as the number (Count) of restarts is not greater than 12 **(En)**.

---

$^1$ **ne** set to unity (see §4.1)
$^2$ Emboldened text refers to input
$^3$ Convergence refers to both neutron and precursor densities

**(A/B/Cn/Dy/Ey/F)** *Convergence of TS?/Last l?/Count > 12?*

If after 12 restarts of Loop **(B)**, not all TS have converged at $t_j$ **(Ey)**, a manual restart **(F)** is called for and the program stops. This means the TS in the partition loop did not all converge at edit $t_j$ after dividing the edit sub-interval into $2^{12}$ sub-intervals. Further interval partitioning is therefore required.

**(A/B/Cy/Gn/Hn/B)** *Convergence of TS?/Convergence of densities?/Last l?*

If the TS has converged for some $l$ in [$l1,l2$] of Loop **(B)**, it is an approximation for the edit at $t_j$ for the $l^{th}$ partition. This adds to the sequence of solutions in $l$ for which we also seek convergence. We then ask if the sequence at $t_j$ has converged. If not **(Gn)**, we check for the last $l$ of the Loop **(B)**. If not **(Hn)**, $l$ is incremented as **(B)** loop continues to build the sequence of solutions at $t_j$.

**(A/B/Cy/Gn/Hy/Iny/B)** *Convergence of TS?/Convergence of densities?/Last l?/Count < > 12?*

If at the last $l$ **(Hy)** and have not exceeded 12 attempts **(In)**, Loop **(B)** restarts with the original edit interval halved **(Gn/Hy/In/B)**. After 12 restarts, a manual restart **(Iy/F)** is required and the program stops.

**(A/B/Cy/Gy/Jn/A)** *Convergence of TS?/Convergence of the densities?/Last j?*

If the TS has converged at some $l$ **(Cy)**, and the sequence of solutions for the densities has converged **(Gy)**, we check for the last $j$. If true **(Jy)**, we are done. If not **(Jn)**, $j$ is incremented and the flow continues through Loops **(A)** and **(B)** until $j$ is $J$.

The manual restart **(F)** is explained in §4.1.

### 3.3 Sequence Convergence

As noted in the algorithmic flow chart, there are two sequence convergences− one for the partial sums of the TS, say for the neutron density at partition $l$,

$$N_j\left(t_j;l\right) = \lim_{N\to\infty}\sum_{n=0}^{N} N_{n,j-1}\left(l\right)\left(t_j - t_{j-1}\right)^n; \tag{7a}$$

and the second for the neutron density at edit $t_j$ converging over the partitions $l$

$$N_j\left(t_j\right) = \lim_{l\to\infty} N_j\left(t_j;l\right). \tag{7b}$$

The means of convergence of Eqs(7) is convergence term-by-term, called natural or sequential convergence. In addition, we apply W-e convergence acceleration to extrapolate the sequences to their limits. W-e acceleration is a non-linear acceleration assuming an extrapolated error series representation in terms of the sequence elements with unknown coefficients. A system of equations is constructed and solved with an extra equation for the approximation of the limit. In principle, the approximation should improve with the number of elements in the error series. Since the initial sequence elements are usually inaccurate, a window of elements for acceleration, nominally consisting of five elements, advances through the sequence list. Note that unlike Richardsons extrapolation [27], there is no regularity to the error series required for W-e making it virtually universal but possibly ineffective. We apply W-e acceleration to the sequences of TS partial sums of Eq(7a) and to the partitioned sequence solutions of Eq(7b) at each edit. Finally, the convergence mode, W-e or natural, whichever results in the least relative error between the last two sequence elements (called the "engineering" error estimate) becomes the sequence limit.

## 4. NUMERICAL CONFIRMATION OF CATS ALGORITHM
### 4.1 Input Parameters and Convergence, "Wrangling" and Agreement
To ensure a meaningful comparison, one compares the DPCATS algorithm (called CATS) to the quadruple precision QPBEFD algorithm (called BEFD). This section describes how one compares the two benchmark solutions. In particular, on what basis is the comparison and to what extent does CATS input adjustment play a role in providing a reliable precision assessment and why?

The following table gives the nominal input parameter list (NIPL) with suggested values for the CATS algorithm:

**Table A. Nominal Input Parameters List (NIPL) for CATS.**

| Parameter | Suggested Value | Description |
|---|---|---|
| **err** | $10^{-12}$-$10^{-15}$ | Desired range relative error |
| **ne** | 1 | Further partition of edit intervals in Loop (**B**) |
| **ne1** | 1 | One time partition of original edit intervals |
| **$l1/l2$** | 4/11,12 | Initial/final partition in Loop (**B**) [$l1 \le l \le l2$] |
| **K** | 15 | Maximum number of terms for TS convergence |
| **m1** | 0 to 3 | TS over convergence index (OCI) (**err/$10^{m1}$**) |
| **n0** | 5 | Number of sequence elements in W-e window |

Parameter **err** is the relative error to determine convergence of the TS partial sums and the solution at each edit over the $l$-partitions. We define relative error as the difference between the last two sequence elements over the most recent element, which is an "engineering" estimate of Cauchy convergence. For example, if we expect 10 places, set the relative error is set in the range of $10^{-12}$ for DP arithmetic. A lower value could possibly incur round off error, though we will see several examples converge with a lower **err**. Several cases use near the lower limit of the allowable range.

Parameters **ne1** and **ne** further divide the edit intervals for additional time step reduction. Parameter **ne1** partitions the original edit intervals into **ne1** partitions,

$$\Delta t_j \rightarrow \frac{\Delta t_j}{\textbf{ne1}} \tag{8}$$

but only once at the start of the calculation, thus reducing the time step to ensure TS convergence within **K** terms. In addition, **ne1** values make up the plots. As noted in the flow chart [Fig. 2] for loop (**B/C/D/E/B**), if the TS has not converged at a given partition, say $l$-1 in Loop (**B**), the next partition interval $h_l$ [Eq(7)] is partitioned by a factor 2*$ne$

$$h_l \rightarrow h_l / (2*ne) = \frac{1}{2*\textbf{ne}} \frac{\Delta t_j}{2^l} \tag{9}$$

and Loop (**B**) restarts. With partition adjustment, the intervals can be made as small as necessary to give efficient TS convergence. In addition, partition adjustments allow numerical "wrangling" (to coin a word from cowboy culture) to be addressed.

$l1$ denotes where the Loop (**B**) partition begins which is important for TS convergence. The larger $l1$ the smaller the time step and faster convergence. Partition $l2$, which nominally can be either 11 or 12, ends the solution sequence for each edit. The larger $l2$ the more elements in the sequence increasing the potential for convergence without further time step reduction, *i.e.*, avoidance of loop (**B/C/D/E/B**) shown in Fig. 2, but requiring more CPU.

The number of terms for TS to converge is limited to **K** for the neutron and precursor densities, which is nominally 15 in current comparisons. If not converged within **K** terms, the TS is declared failed and ignored, the partition advances to $l$+1 and the calculation of Loop (**B**) continues. The same is true if convergence is to a negative density. Referring to the flow chart of Fig. 2, failure to converge to a positive value at or before partition $l2$ (**En**) forces further partitions (by factor 2***ne**, see Eq(9)) and Loop (**B**) continues. Any combination of TS failures can occur in [$l1$, $l2$] 12 times before a manual adjustment is necessary (**Ey/F**). One then proceeds by choosing from the following adjustments and restarting:

**(a)** Increase the initial number of partitions (*i.e.*, **ne1**) to reduce the time step in Loop (**B**) in order to induce TS convergence.

**(b)** Increase *l1* to further reduce the step or increase *l2* to extend partitions to produce more sequence elements.

**(c)** Increase **K** to ease the TS convergence limitation.

**(d)** Reduce or increase the over convergence index (OCI, *i.e.*, **m1**) to converge the TS in fewer terms or with higher precision. To be considered nominal, **m1** is 0 to 3.

**(e)** Increase the W-e window **n0** to provide a better opportunity for W-e convergence. **n0** should be no larger than 15.

**(f)** Increase or decrease the desired error **err.**

Thus, selections of adjustments are available to enable TS convergence and numerical "wrangling" for full agreement with BEFD if possible.

CATS solutions are not unique with regard to the choice of input parameters because of the considerable number of input parameters (7) with each choice defining a different path to sequence convergence. Since extreme benchmarks live between the $9^{th}$ and $12^{th}$ digits, we are comparing solutions at the margin of DP computational precision, which practically is no more than twelve digits. Thus, the choice of the IPL has a significant impact on the last digit. Here is where "wrangling" comes in. Since we are in the unique position of knowing the 12- place benchmark from BEFD, one could simply adjust or "wrangle" the partitions through the IPL to try to eventually match the known benchmark by trial and error. But what value is such an exercise with regard to verification of the CATS algorithm since, except in this rare situation, we never have the true benchmark to compare to? The best outcome would be, on convergence, all 12 places match (within the NIPL ranges of *l2*, **m1** and **err)** and CATS would be verified to 12 places. Chances of this occurring are small. Wrangling the IPL therefore offers the possibility of verifying a number of places, say 10, which we know to be true, thus providing a measure of precision of the CATS algorithm. The least amount of wrangling, the higher the confidence in the precision. If no benchmark like BEFD is available, then wrangling is the same as a sensitivity study, which is common sense verification for any solution. The precision would then be how close two solutions are with respect to number of places when say, **err** or **m1** are varied. For the purpose of our investigation however, we use wrangling in a controlled way to addresses our curiosity regarding exactly how close the two algorithms actually are. Thus, wrangling does play a role in CATS verification at several levels.

To limit excessive wrangling possibly leading nowhere and to provide some consistency, we suggest the following wrangling guidelines based on the application of the emboldened values (NIPL) of Table A. If agreement is less than ten or eleven places:

**(a)** Change **m1** from 0 to 3 with *l2* = 11.
**(b)** Still no agreement, increase *l2* to 12 for better TS convergence and again try **m1** from 0 to 3.
**(c)** Still no agreement, reduce relative error **err** to $10^{-13}$, $10^{-14}$ or $10^{-15}$. Chances are convergence will be lost when **err** is too low.
**(d)** Still no agreement, set *l2* to 14, 15 or 16 and **m1** = 2 or 3.
**(e)** Still no agreement, increase **ne1** to 16 or **ne** to 16 or both for **(c)** or **(d)**.
**(f)** Set **ne1** to 100 in **(e)**.

For each benchmark, the outcome, with or without wrangling, is included as a table of BEFD results with an overlay of CATS results and the differences marked. In this way, the 12-place BEFD benchmark appears and the precision of CATS is evident. An accompanying narrative gives the details of the wrangling required. Two categories of Wrangling exist, minor and major. Minor wrangling includes steps (**a**,**b**,**c**,**d**), which would naturally be part of a sensitivity study of any solution. Major wrangling includes steps (**e**,**f**), which are not obvious and require guessing and trial and error. Minor wrangling could conceivably be applied to all CATS benchmarks as a sensitivity study, but major wrangling is for a known benchmark thus requiring, in our case, BEFD. We record all input lists in Table B of Appendix B including computational time.

Several comments are in order. While 10- place agreement emerges most readily, 11 and 12 places may not. Here is where one wrangles just to see if higher agreement is indeed possible. However, noting that 10 places accomplishes our original extreme benchmark goal, we limit excessive wrangling usually to several discrepancies in the twelfth place. The premise is that a 10-place CATS benchmark is sufficient since it is still high precision in the PKE literature; but if we can establish more places, all the better. Note however, that we always publish the 12- place BEFD benchmark as a reference.

Since we are creating a benchmark for a given set of edits (usually from 5 to 10), the precision of any benchmark depends upon the choice of edits because they determine the partitions. So there is a degree of uncontrolled uncertainty associated with establishing a high precision benchmark, which is unavoidable. However, experience has shown that 10 places are relatively independent of variation of the choice of the IPL and edits as will become evident.

Finally, a third comparison option is available and that is quadruple precision CATS (QPCATS); however, CPU will be greatly increased. Nevertheless, QP is a last resort for verification, which we apply on three occasions in this investigation.

## 4.2 An Extreme Benchmark Comparison

To show that the CATS algorithm indeed gives a reliable benchmark, we consider a step reactivity insertion of $0.5 in Fast Reactor I (FRI) in the extreme. Reactor kinetic parameters for all benchmarks in our study are in the tables of Appendix A with FRI in Table A.I. The first appearance of this benchmark seems to be in Nobrega [9] and ever since has served as a standard for comparison for newly developed numerical RKE algorithms. It is well known for step reactivities there exists an analytical solution via diagonalization of the Jacobian matrix to solve the scalar ODEs on the diagonal coupled through eigenvectors. Since the analytical solution has appeared in texts [see Ref. 28, for example], we skip its derivation.

The existence of the BEFD algorithm [2] puts us in a unique situation regarding comparisons. First published to precision better than one unit in the ninth place, the BEFD algorithm is a fundamental finite difference solution with iteration for non-linear reactivity feedback. The BEFD algorithm also includes sequence convergence with acceleration on an adaptive mesh to mitigate propagation error. Recently, in a a relatively thorough investigation, the BEFD algorithm was compared to the physics informed neural net (PINN) precise formulation X-TFC [29] to, in some cases, 12-places. The PINN comparison gives confidence in the BEFD benchmark for what follows. However, to ensure a proper comparison, we execute BEFD in quadruple precision to be certain of at least 12-places of precision when compared to CATS in double precision.

For all comparisons, the number of digits minus one in agreement (number of places), primarily for the neutron density at a given time edit, is the benchmarking measure. Each case includes a table of only the BEFD benchmark results for 10, 11 and 12 places, expected to be precise including the last digit quoted. The discrepant digits in comparison to CATS will be emboldened and overlaid onto the table. An underscored digit, such as $\underline{\mathbf{X}}$, indicates the CATS digit is below BEFD; and if underscored, such as $\underline{\mathbf{X}}$, CATS is above BEFD. In either case, the discrepancy is a single unit. If the digit is only highlighted, then the discrepancy is more than one unit either above or below BEFD.

Table 1a represents the results of an extreme precision example to 23 places for step reactivity insertion with both BEFD and CATS in QP and including the analytical solution in QP by diagonalization. This is the first or three instances of CATS executed in QP. One finds perfect agreement to all places for all three solutions, which gives confidence in the BEFD benchmark for the CATS comparisons to follow.

**Table 1a. Extreme precision demonstration: Step insertion into FRI**
**($\rho_0$ = $0.5, $\Lambda = 10^{-7}$)**

| t(s) | QPBEFD/QPCATS/QPAnalytical |
|---|---|
| 0.0000E+00 | 1.00000000000000000000000E+00 |
| 1.0000E -01 | 2.07531716248122033152682E+00 |
| 1.0000E+00 | 2.65585295935837676143711E+00 |
| 1.0000E+01 | 1.27465396678628924693564E+01 |
| 1.0000E+02 | 1.31287276940349626158236E+07 |

With BEFD established as a credible benchmark, at least for step insertions, in the following demonstrations, we compare DPCATS to QPBEFD (called CATS and BEFD respectively) anticipating at least ten or more place agreement. As will be shown, because of variability associated with the IPL, in particular partitioning, the eleventh or twelfth place can require numerical wrangling to achieve full agreement—especially for the twelfth place. We now continue toward our goal of a minimum of 10 places of agreement for a common NIPL including only minor wrangling for our examples and possibly full agreement with major wrangling.

### 4.3 Further Verification

Table 1b is a continuation of the initial demonstration for the IPL of Table A to establish a 10- place benchmark. As observed, both benchmarks agree to all 12 digits (since no digits are emboldened). Column five includes the values of the original benchmark to show the contrast with what one considered a benchmark in the past. The last column shows that (solution) convergence at the edits is split between W-e acceleration (2) and sequential (1). TS convergence is almost entirely from W-e acceleration.

**Table 1b. Neutron density confirmation: Step reactivity into FRI**
**($\rho_0$ = $0.5, $\Lambda = 10^{-7}$)**

| t(s) | BEFD/CATS(10) | BEFD/CATS(11) | BEFD/CATS(12) | Reference[9] | conv |
|---|---|---|---|---|---|
| 0.0000E+00 | 1.0000000000E+00 | 1.00000000000E+00 | 1.000000000000E+00 | 1.000000E+00 | |
| 1.0000E -01 | 2.0753171625E+00 | 2.07531716248E+00 | 2.075317162481E+00 | 2.075317E+00 | 1 |
| 1.0000E+00 | 2.6558529594E+00 | 2.65585295936E+00 | 2.655852959358E+00 | 2.655853E+00 | 1 |
| 1.0000E+01 | 1.2746539668E+01 | 1.27465396679E+01 | 1.274653966786E+01 | 1.274654E+01 | 2 |
| 1.0000E+02 | 1.3128727694E+07 | 1.31287276940E+07 | 1.312872769403E+07 | ---- | 2 |

Table 1c compares the precursor densities to 11 places. For the NIPL there is only one discrepancy in the 11th place in delayed group 2 at $t = 10s$. Seems no amount of wrangling eliminates this discrepancy without causing discrepancies at other entries.

**Table 1c**. **Precursor density confirmation**: Step reactivity into FRI
($\rho_0$ = $0.5, $\Lambda = 10^{-7}$)

| $t(s)/m$ | 1 | 2 | 3 | 4 | 5 | 6 |
|---|---|---|---|---|---|---|
| 0.0000E+00 | 1.29612403101E+05 | 3.96141479100E+05 | 7.09253731343E+04 | 4.35951661631E+04 | 3.59841269841E+03 | 4.79750778816E+02 |
| 1.0000E- 01 | 1.29785770834E+05 | 3.97417778603E+05 | 7.19049645570E+04 | 4.50681476487E+04 | 4.04075603004E+03 | 6.16699564464E+02 |
| 1.0000E+00 | 1.31839871541E+05 | 4.12418201067E+05 | 8.29098959659E+04 | 6.02777250697E+04 | 7.22324746594E+03 | 1.16488836072E+03 |
| 1.0000E+01 | 2.13283235626E+05 | 9.77487790540E+05 | 4.04343149675E+05 | 3.71865336497E+05 | 4.07151646027E+04 | 5.82683889600E+03 |
| 1.0000E+02 | 1.31850995210E+11 | 8.75792669859E+11 | 4.33872945342E+11 | 3.90947865256E+11 | 4.21097025638E+10 | 6.01091977254E+09 |

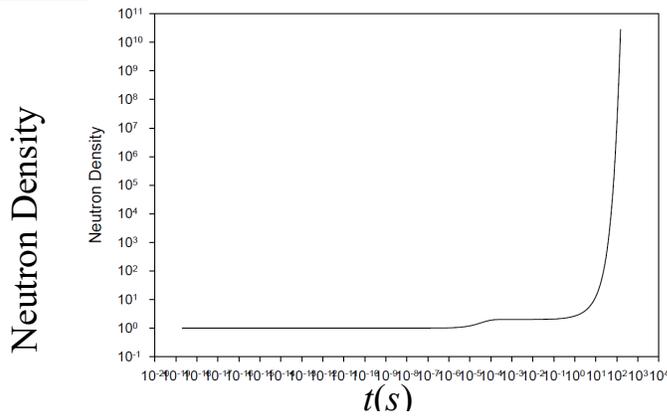

Fig. 3. Decadal time variation of the density step reactivity into FRI.

**Table 1d. Decadal Variation: Step reactivity into FRI**
($\rho_0$ = $0.5, $\Lambda = 10^{-7}$)

| $t(s)$ | BEFD/CATS(10) | BEFD/CATS(11) | BEFD/CATS(12) | $conv$ |
|---|---|---|---|---|
| 1.0000E- 15 | 1.0000000000E+00 | 1.00000000002E+00 | 1.000000000022E+00 | |
| 1.0000E- 14 | 1.0000000002E+00 | 1.00000000022E+00 | 1.000000000220E+00 | 2 |
| 1.0000E- 13 | 1.0000000022E+00 | 1.00000000220E+00 | 1.000000002200E+00 | 1 |
| 1.0000E- 12 | 1.0000000220E+00 | 1.00000002200E+00 | 1.000000022000E+00 | 2 |
| 1.0000E- 11 | 1.0000002200E+00 | 1.00000022000E+00 | 1.000000220000E+00 | 2 |
| 1.0000E- 10 | 1.0000022000E+00 | 1.00000220000E+00 | 1.000002199998E+00 | 2 |
| 1.0000E- 09 | 1.0000219998E+00 | 1.00002199976E+00 | 1.000021999758E+00 | 1 |
| 1.0000E- 08 | 1.0002199758E+00 | 1.00021997580E+00 | 1.000219975802E+00 | 2 |
| 1.0000E- 07 | 1.0021975818E+00 | 1.00219758177E+00 | 1.002197581774E+00 | 2 |
| 1.0000E- 06 | 1.0217597650E+00 | 1.02175976501E+00 | 1.021759765011E+00 | 2 |
| 1.0000E- 05 | 1.1974812583E+00 | 1.19748125831E+00 | 1.197481258310E+00 | 2 |
| 1.0000E- 04 | 1.8892203637E+00 | 1.88922036370E+00 | 1.889220363704E+00 | 1 |
| 1.0000E- 03 | 2.0007068176E+00 | 2.00070681763E+00 | 2.000706817634E+00 | 1 |
| 1.0000E- 02 | 2.0076811958E+00 | 2.00768119583E+00 | 2.007681195831E+00 | 1 |
| 1.0000E- 01 | 2.0753171625E+00 | 2.07531716248E+00 | 2.075317162481E+00 | 1 |
| 1.0000E+00 | 2.6558529594E+00 | 2.65585295936E+00 | 2.655852959358E+00 | 2 |
| 1.0000E+01 | 1.2746539668E+01 | 1.27465396679E+01 | 1.274653966786E+01 | 2 |
| 1.0000E+02 | 1.3128727694E+07 | 1.31287276940E+07 | 1.312872769403E+07 | 2 |

However, QPCATS agrees with BEFD at 12 times the CPU of DPCATS and is the second instance where QP is use to show agreement.

Figure 3 displays the decadal neutron density trace for a $0.5 step showing the initiation of the delayed neutron contribution at early time and the prompt neutrons at late time. Table 1d gives a decadal benchmark over 18 temporal decades as confirmed by BEFD to all digits quoted. The primary mode of convergence is W-e acceleration.

With the benchmarking procedure now in place, we are ready to demonstrate the CATS high precision benchmark on a variety of standard insertions.

## 5. SELECTED DEMONSTRATIONS OF PRESCRIBED REACTIVITY INSERTIONS

The following demonstrations are representative of cases found in the literature including standard prescribed reactivity insertions covering negative step, ramp and sinusoidal. For all cases, a table of 10-, 11- and 12- place benchmarks is presented for the neutron density from the BEFD benchmark determined in QP to 14 or 15 places with CATS results overlaid (see §4.2).

### 5.1. Negative Step Insertions

We now consider Thermal Reactor I (TRI) with parameters from Table A.II of Appendix A for negative step insertions of $-$0.5 and $-$1. We see perfect agreement in Tables 2a,b between the two algorithms.

**Tables 2a,b. Negative Step:$-$0.5/$-$1 Step reactivities into TRI**
**($\rho_0$ = $0.5/$1, $\Lambda$ = $5 \times 10^{-4}$)**

| $\rho$ | $t(s)$ | BEFD/CATS(10) | BEFD/CATS(11) | BEFD/CATS(12) | conv |
|---|---|---|---|---|---|
| (a) $-$0.5 | 0.0000E+00 | 1.0000000000E+00 | 1.00000000000E+00 | 1.000000000000E+00 | |
| | 1.0000E-01 | 6.9892522557E-01 | 6.98925225571E-01 | 6.989252255707E-01 | 2 |
| | 1.0000E+00 | 6.0705356561E-01 | 6.07053565606E-01 | 6.070535656062E-01 | 2 |
| | 5.0000E+00 | 4.8255302992E-01 | 4.82553029922E-01 | 4.825530299222E-01 | 2 |
| | 1.0000E+01 | 3.9607769072E-01 | 3.96077690715E-01 | 3.960776907155E-01 | 2 |
| | 1.0000E+02 | 7.1582854439E-02 | 7.15828544386E-02 | 7.158285443859E-02 | 1 |
| | 2.0000E+02 | 1.6934791659E-02 | 1.69347916587E-02 | 1.693479165875E-02 | 2 |
| (b) $-$1.0 | 0.0000E+00 | 1.0000000000E+00 | 1.00000000000E+00 | 1.000000000000E+00 | |
| | 1.0000E-01 | 5.2056428661E-01 | 5.20564286609E-01 | 5.205642866088E-01 | 2 |
| | 1.0000E+00 | 4.3333344530E-01 | 4.33333445301E-01 | 4.333334453006E-01 | 1 |
| | 5.0000E+00 | 3.1067776843E-01 | 3.10677768428E-01 | 3.106777684278E-01 | 1 |
| | 1.0000E+01 | 2.3611065079E-01 | 2.36110650789E-01 | 2.361106507888E-01 | 2 |
| | 1.0000E+02 | 2.8667642454E-02 | 2.86676424545E-02 | 2.866764245448E-02 | 2 |
| | 2.0000E+02 | 5.5905967933E-03 | 5.59059679335E-03 | 5.590596793349E-03 | 2 |

## 5.2. Ramp Insertions

We next consider ramp insertions of $0.1/s$ and $1/s$ in Thermal Reactor II whose kinetic parameters are in Table A.III of Appendix A. The PKEs for a reactivity ramp insertion does not have an analytical solution and can reach large densities quickly as shown in Table 3a for the first ramp. Again, we find perfect agreement.

### Table 3a. Ramp Reactivity: $0.1/s$ ramp into TRII
### ($\Lambda = 5x10^{-4}$)

| $t(s)$ | BEFD/CATS(10) | BEFD/CATS(11) | BEFD/CATS(12) | conv |
|--------|---------------|---------------|---------------|------|
| 0.0000E+00 | 1.0000000000E+00 | 1.00000000000E+00 | 1.000000000000E+00 | |
| 2.0000E+00 | 1.3382000500E+00 | 1.33820005005E+00 | 1.338200050049E+00 | 2 |
| 4.0000E+00 | 2.2284418968E+00 | 2.22844189681E+00 | 2.228441896810E+00 | 1 |
| 6.0000E+00 | 5.5820524487E+00 | 5.58205244867E+00 | 5.582052448674E+00 | 1 |
| 8.0000E+00 | 4.2786295731E+01 | 4.27862957311E+01 | 4.278629573112E+01 | 1 |
| 1.0000E+01 | 4.5116362391E+05 | 4.51163623909E+05 | 4.511636239090E+05 | 2 |
| 1.1000E+01 | 1.7922136073E+16 | 1.79221360734E+16 | 1.792213607343E+16 | 1 |

### Table 3b. Ramp Reactivity: $1/s$ ramp into FRI
### ($\Lambda = 5x10^{-4}$, err = $8x10^{-15}$)

| $t(s)$ | BEFD/CATS(11) | BEFD/CATS(11) | BEFD/CATS(12) | conv |
|--------|---------------|---------------|---------------|------|
| 0.0000E+00 | 1.0000000000E+00 | 1.00000000000E+00 | 1.000000000000E+00 | |
| 1.0000E- 02 | 1.0100971112E+00 | 1.01009711118E+00 | 1.010097111175E+00 | 2 |
| 1.0000E- 01 | 1.1133201123E+00 | 1.11332011226E+00 | 1.113320112264E+00 | 2 |
| 2.0000E- 01 | 1.2605599250E+00 | 1.26055992496E+00 | 1.260559924957E+00 | 2 |
| 5.0000E- 01 | 2.1364091074E+00 | 2.13640910738E+00 | 2.136409107379E+00 | 2 |
| 1.0000E+00 | 1.2078141972E+03 | 1.20781419722E+03 | 1.207814197221E+03 | 1 |
| 1.1000E+00 | 3.2575933547E+99 | 3.25759335471E+99 | 3.257593354712E+99 | 2 |
| 1.1500E+00 | 1.0289753595+219 | 1.02897535952+219 | 1.028975359518+219 | 1 |

For the second ramp insertion of $1/s$ into FRI with the comparison shown in Table 3b for the nominal case with **err** set to $8x10^{-15}$ and **m1** set to 2, we see agreement for 10 and 11 places, and one entry multiple units off for 12 places. Further, if we continue wrangling and let **ne** = **ne1** = 16 and **l1** = 2, which are not obvious changes, we get perfect agreement. Thus, minor wrangling is unnecessary for a 10- or 11-place benchmark; but to reproduce BEFD perfectly, major wrangling is necessary.

## 5.3 Zig- Zag Insertion

We now come to one of the more exotic insertions, the Zig- Zag insertion into TRI (Table A.II). The Zig- Zag insertion consist of four legs of up and down insertions as follows:

$$\rho(t) = \begin{cases} 0.0075t, & 0 \le t \le 0.5 \\ -0.0075(t-0.5)+0.00375, & 0.5 \le t \le 1 \\ 0.0075(t-1), & 1 \le t \le 1.5 \\ 0.00375, & 1.5 \le t \end{cases} \tag{10}$$

giving a piecewise continuous insertion, but not piecewise smooth. For this reason, each leg is viewed as a separate insertion so the TS solution is applied piecewise. As observed in Table 4 perfect agreement is achieved for the NIPL with **m1** = 2. Also note the majority of edits converged by W-e acceleration.

The Zig- Zag example shows by segmenting the reactivity, any set of piecewise continuous reactivity insertions is permissible.

### Table 4. Zig- Zag Insertion into TRI
### ($\Lambda = 5x10^{-4}$)

| $t(s)$ | BEFD/CATS(10) | BEFD/CATS(11) | BEFD/CATS(12) | conv |
|--------|---------------|---------------|---------------|------|
| 0.0000E- 01 | 1.0000000000E+00 | 1.00000000000E+00 | 1.000000000000E+00 | |
| 5.0000E- 01 | 1.7214224221E+00 | 1.72142242208E+00 | 1.721422422085E+00 | 2 |
| 1.0000E+00 | 1.2111274148E+00 | 1.21112741483E+00 | 1.211127414825E+00 | 1 |
| 1.5000E+00 | 1.8922261404E+00 | 1.89222614039E+00 | 1.892226140394E+00 | 2 |
| 2.0000E+00 | 2.5216005300E+00 | 2.52160053000E+00 | 2.521600530004E+00 | 2 |
| 1.0000E+01 | 1.2047105355E+01 | 1.20471053548E+01 | 1.204710535483E+01 | 2 |
| 1.0000E+02 | 6.8155568885E+07 | 6.81555688848E+07 | 6.815556888477E+07 | 2 |

### 5.4 Sinusoidal Insertions

Next, we consider 1- delayed group sinusoidal insertion into Fast Reactor II (FRII) with kinetic parameters

$$\begin{aligned} \beta &= 0.0079 \\ \lambda &= 0.077 \\ \Lambda &= 1x10^{-7} s. \end{aligned} \tag{11}$$

For this fast transient, the reactivity varies as

$$\rho(t) = \rho_0 \sin(\gamma t). \tag{12a}$$

The TS coefficients are

$$\rho_{n,j} = (-1)^{[n/2]} \frac{\gamma^n}{n!} \rho_0 \begin{cases} \sin(\gamma t_j), n \text{ even} \\ \cos(\gamma t_j), n \text{ odd}, \end{cases} \tag{12b}$$

or more concisely

$$\rho_{n,j} = \frac{\gamma^n}{n!} \rho_0 \left[ \cos\left(n\pi/2\right)\sin\left(\gamma t_j\right) + \sin\left(n\pi/2\right)\cos\left(\gamma t_j\right) \right], \qquad (12c)$$

which simplifies to

$$\rho_{n,j} = \frac{\gamma^n}{n!} \rho_0 \sin\left(\gamma t_j + n\pi/2\right), \qquad (13)$$

where $\gamma \equiv \pi/50$ and $\rho_0 = \$0.005333$.

Figure 4 is a plot of the neutron density variation and the benchmark comparison is in Table 5a. With an OCI of **m1** = 3, we find convergence to 10 and 11 places as well as 12 places below $t = 10s$. The CPU time for Table 5a was exceptionally long at $107s$. Of course, we have at least 11 places of precision; but if only 9 places were called for, the CPU would be $30s$ less.

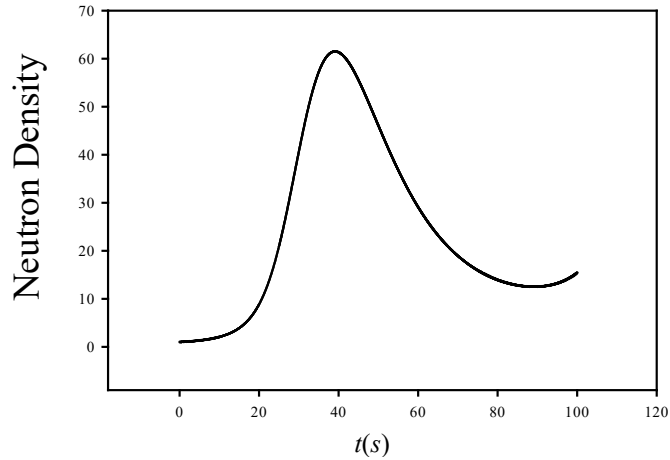

Fig. 4. Neutron density for 1 delayed group sinusoidal reactivity insertion into FRII.

**Table 5a. 1- delayed group sinusoidal reactivity insertion into FRII**
($\rho_0 = \$0.005333,\ \gamma \equiv \pi/50,\ \Lambda = 10^{-7}$)

| $t(s)$ | BEFD/CATS(10) | BEFD/CATS(11) | BEFD/CATS(12) | conv |
|---|---|---|---|---|
| 0.0000E+00 | 1.0000000000E+00 | 1.00000000000E+00 | 1.000000000000E+00 | |
| 1.0000E- 02 | 1.0004239597E+00 | 1.00042395973E+00 | 1.000423959732E+00 | 1 |
| 1.0000E- 01 | 1.0042754809E+00 | 1.00427548091E+00 | 1.004275480913E+00 | 2 |
| 1.0000E+00 | 1.0460190642E+00 | 1.04601906418E+00 | 1.046019064184E+00 | 1 |
| 1.0000E+01 | 2.0652644823E+00 | 2.06526448228E+00 | 2.065264482276E+00 | 1 |
| 1.0000E+02 | 1.5440238789E+01 | 1.54402387889E+01 | 1.544023878889E+01 | 1 |

For the second sinusoidal variation, we consider a 6- delayed group Thermal Reactor I (Table A. II) version, where $\gamma \equiv \pi / 150$ and $\rho_0 = \$0.003233$ with results shown in Table 5b. One observes 12- place precision, when $l1 = 5$ and $m1 = 1$, but for one unit in the last entry.

**Table 5b. 6- delayed group sinusoidal reactivity insertion into TRI**
$$\left( \rho_0 = \$0.003233, \ \gamma \equiv \pi / 150 \right)$$

| $t(s)$ | BEFD/CATS(10) | BEFD/CATS(11) | BEFD/CATS(12) | conv |
|--------|---------------|---------------|---------------|------|
| 0.0000E+00 | 1.0000000000E+00 | 1.00000000000E+00 | 1.000000000000E+00 | |
| 1.0000E- 02 | 1.0000064453E+00 | 1.00000644531E+00 | 1.000006445312E+00 | 2 |
| 1.0000E- 01 | 1.0004374069E+00 | 1.00043740686E+00 | 1.000437406860E+00 | 1 |
| 1.0000E+00 | 1.0096671690E+00 | 1.00966716902E+00 | 1.009667169017E+00 | 2 |
| 1.0000E+01 | 1.2039760355E+00 | 1.20397603551E+00 | 1.203976035506E+00 | 2 |
| 1.0000E+02 | 3.2103073547E+03 | 3.21030735465E+03 | 3.210307354653E+03 | 2 |

# 6. SELECTED DEMONSTRATIONS OF NON-LINEAR ADIABATIC DOPPLER AND DENSITY- DEPENDENT INSERTIONS

We now are concerned with adiabatic Doppler feedback reactivity of the form

$$\rho(t) = \rho_0(t) - B \int_0^t dt' N(t'),$$ (14a)

where $\rho_0(t)$ is prescribed and $B$ is the given temperature coefficient of reactivity. The second term is the Doppler integral term adding negative reactivity proportional to reactor heating. By taking the $n^{\text{th}}$ derivative and noting

$$\frac{d^n}{dt^n} \int_0^t dt' N(t') = \frac{d^{n-1}}{dt^{n-1}} \frac{d}{dt} \int_0^t dt' N(t') = N^{(n)}(t),$$ (14b)

the $n^{\text{th}}$ TS coefficient of reactivity at edit $t_j$ is simply

$$\rho_{n,j} = \rho_{0,n,j} - \frac{B}{n} N_{n-1,j}.$$ (15)

## 6.1 Step insertion with Doppler

For a step insertion, $\rho_0$, we have, for $N_{-1,j} \equiv 0$, the following TS coefficient:

$$\rho_{n,j} = \rho_0 \delta_{n,0} - \frac{B}{k} N_{n-1,j}.$$ (16)

Now consider the kinetics parameters of Thermal Reactor IV in Table V with $\Lambda = 5x10^{-5}s$ and $B = 2.5x10^{-6}$/MWs. Figure 5 shows how the Doppler feedback effectively stops transients of step reactivity insertions of \$0.1, to \$2. Tables 6a,b,c give neutron densities for insertions of \$1, 1.5 and \$2, respectively. For the \$1 insertion with **m1** = 1, we observe one discrepancy in the 12th digit by a single unit. When **err** is $10^{-14}$, the discrepancy vanishes. For the \$1.5 insertion with the same NIPL, there are no discrepancies for **m1** set to 2 and **l2** set to 12, respectively.

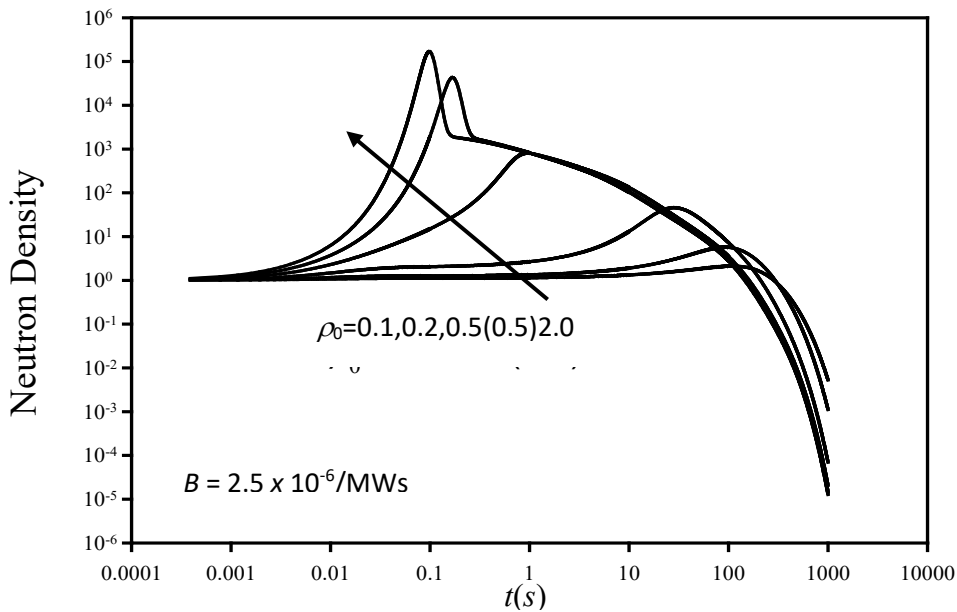

Fig. 5. Neutron density for adiabatic feedback and six step insertions.

## Table 6a. $\rho_0$ = \$1

| $t(s)$ | BEFD/CATS(10) | BEFD/CATS(11) | BEFD/CATS(12) | conv |
|---|---|---|---|---|
| 0.0000E+00 | 1.0000000000E+00 | 1.00000000000E+00 | 1.000000000000E+00 | |
| 1.0000E+00 | 8.0527812545E+02 | 8.05278125449E+02 | 8.052781254488E+02 | 2 |
| 1.0000E+01 | 1.3194585878E+02 | 1.31945858778E+02 | 1.319458587776E+02 | 2 |
| 2.0000E+01 | 5.1675599140E+01 | 5.16755991398E+01 | 5.167559913980E+01 | 2 |
| 3.0000E+01 | 2.8165478123E+01 | 2.81654781231E+01 | 2.816547812308E+01 | 2 |
| 4.0000E+01 | 1.8141812575E+01 | 1.81418125753E+01 | 1.814181257533E+01 | 2 |
| 5.0000E+01 | 1.2776961432E+01 | 1.27769614317E+01 | 1.277696143166E+01 | 2 |
| 6.0000E+01 | 9.4732419528E+00 | 9.47324195281E+00 | 9.473241952811E+00 | 2 |
| 7.0000E+01 | 7.2433002064E+00 | 7.24330020643E+00 | 7.243300206432E+00 | 2 |
| 8.0000E+01 | 5.6454283755E+00 | 5.64542837546E+00 | 5.645428375455E+00 | 2 |
| 9.0000E+01 | 4.4561837084E+00 | 4.45618370844E+00 | 4.456183708438E+00 | 2 |
| 1.0000E+02 | 3.5496012141E+00 | 3.54960121408E+00 | 3.549601214084E+00 | 2 |

**Table 6b. $\rho_0 = \$1.5$**

| | | | | |
|---|---|---|---|---|
| 0.0000E+00 | 1.0000000000E+00 | 1.00000000000E+00 | 1.000000000000E+00 | |
| 1.0000E+00 | 8.1094972780E+02 | 8.10949727799E+02 | 8.109497277989E+02 | 1 |
| 1.0000E+01 | 1.0787694512E+02 | 1.07876945123E+02 | 1.078769451229E+02 | 1 |
| 2.0000E+01 | 4.1596896088E+01 | 4.15968960884E+01 | 4.159689608837E+01 | 2 |
| 3.0000E+01 | 2.3296227711E+01 | 2.32962277105E+01 | 2.329622771050E+01 | 2 |
| 4.0000E+01 | 1.5302087318E+01 | 1.53020873175E+01 | 1.530208731751E+01 | 2 |
| 5.0000E+01 | 1.0889353812E+01 | 1.08893538118E+01 | 1.088935381184E+01 | 2 |
| 6.0000E+01 | 8.1005133978E+00 | 8.10051339775E+00 | 8.100513397754E+00 | 1 |
| 7.0000E+01 | 6.1823251167E+00 | 6.18232511674E+00 | 6.182325116740E+00 | 1 |
| 8.0000E+01 | 4.7930388494E+00 | 4.79303884942E+00 | 4.793038849418E+00 | 1 |
| 9.0000E+01 | 3.7554112321E+00 | 3.75541123207E+00 | 3.755411232072E+00 | 2 |
| 1.0000E+02 | 2.9659185490E+00 | 2.96591854900E+00 | 2.965918548995E+00 | 2 |

**Table 6c. $\rho_0 = \$2$**

| | | | | |
|---|---|---|---|---|
| 0.0000E+00 | 1.0000000000E+00 | 1.00000000000E+00 | 1.000000000000E+00 | |
| 1.0000E+00 | 8.1329909944E+02 | 8.13299099435E+02 | 8.132990994355E+02 | 2 |
| 1.0000E+01 | 1.0335691805E+02 | 1.03356918046E+02 | 1.033569180456E+02 | 2 |
| 2.0000E+01 | 3.9134327846E+01 | 3.91343278459E+01 | 3.913432784591E+01 | 1 |
| 3.0000E+01 | 2.2002158526E+01 | 2.20021585264E+01 | 2.200215852640E+01 | 2 |
| 4.0000E+01 | 1.4492870628E+01 | 1.44928706275E+01 | 1.449287062751E+01 | 2 |
| 5.0000E+01 | 1.0318137565E+01 | 1.03181375650E+01 | 1.031813756503E+01 | 2 |
| 6.0000E+01 | 7.6630069808E+00 | 7.66300698080E+00 | 7.663006980800E+00 | 1 |
| 7.0000E+01 | 5.8291748686E+00 | 5.82917486862E+00 | 5.829174868621E+00 | 2 |
| 8.0000E+01 | 4.4992647619E+00 | 4.49926476188E+00 | 4.499264761876E+00 | 2 |
| 9.0000E+01 | 3.5073002243E+00 | 3.50730022434E+00 | 3.507300224338E+00 | 1 |
| 1.0000E+02 | 2.7550331306E+00 | 2.75503313062E+00 | 2.755033130619E+00 | 1 |

We see perfect agreement for the \$2 insertion when **m1** is 2.

It is comforting that the NIPL gives perfect agreement between the benchmarks with only a minor adjustment of the NIPL.

### 6.2 Keepin's [11] Compensated Ramp Insertion

We next consider Keepin's compensated ramp insertion with adiabatic Doppler feedback with

$$
\begin{aligned}
\rho_{0,j} &= 0 \\
\rho_{1,j} &= a \\
\rho_{n,j} &= -\frac{B}{n} N_{n-1,j}, \quad n = 2, 3, ..
\end{aligned}
\tag{17}
$$

and proceed as previously for four ramps rates ($a$) and two Doppler coefficients ($B$).

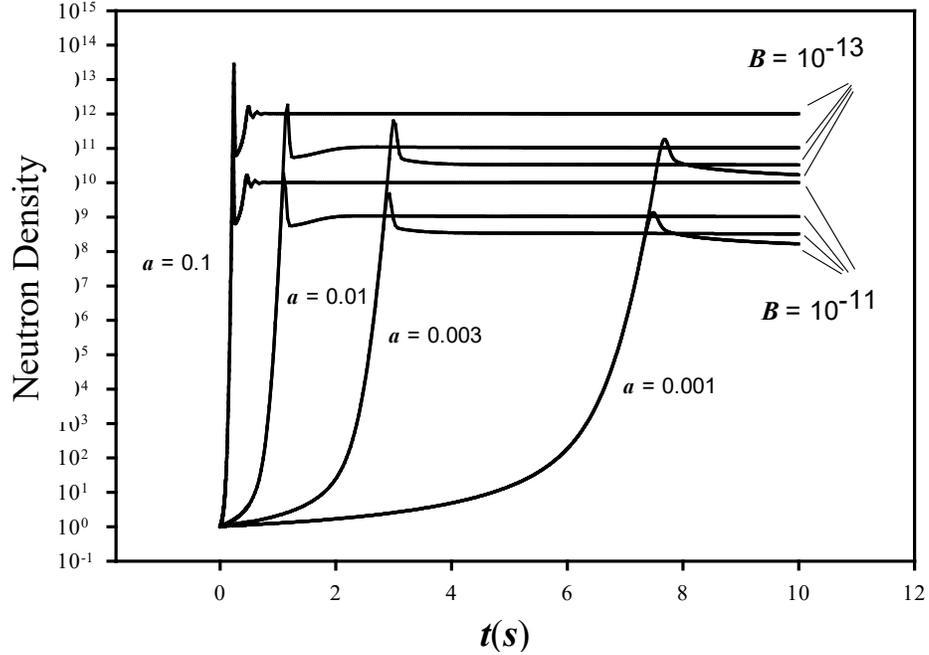

Fig. 6. Compensated ramp insertion as in Keepin [11].

Figure 6 shows the Doppler effect again effectively shutting down the transient after a short but increasing time with decreasing ramp rate initiating the transient. The figure is virtually identical to that found in the classic work of Keepin [11]. Tables 7a –h present a comparison between CATS and BEFD neutron densities for the eight transients. For the NIPL with **m1** = 0, all 11 places agree with only 6 entries for all tables disagreeing in the $12^{th}$ place (not shown). For **m1** = 2 in all but insertions c, and g, where **m1** is 1, all discrepancies vanish except for the second to last entry of case 8 (emboldened), which does vanish if **err** is set to $5x10^{-13}$ as found by trial and error (or minor wrangling).

One should note that more than 70% of the edits (51/72) converge by convergence acceleration in $0.67s$ of CPU time. If we suppress acceleration, the computational time reduces to $0.14s$ but with 3 discrepancies. In addition, if we only allow acceleration, the CPU time is $7s$ with one discrepancy. Curiously, all three calculations with acceleration, with no acceleration and for a combination of acceleration and no acceleration have the same discrepant entry in Table e. Including acceleration does increase computational time; but, nevertheless, enables the choice of either natural or accelerated convergence, which seems to have a significant time advantage over acceleration only. Most importantly, including acceleration increases confidence in the computation as it provides an independent verification.

## Tables 7a-h. Neutron Densities for ramps into TRIII with Doppler)

### (a) $a = 0.1$  $B = 10^{-11}$

| $t(s)$ | BEFD/CATS(10) | BEFD/CATS(11) | BEFD/CATS(12) | conv |
|---|---|---|---|---|
| 0.0000E+00 | 1.0000000000E+00 | 1.00000000000E+00 | 1.000000000000E+00 | |
| 1.0000E-01 | 2.4733658212E+01 | 2.47336582123E+01 | 2.473365821230E+01 | 2 |
| 5.0000E-01 | 9.9498932412E+09 | 9.94989324117E+09 | 9.949893241175E+09 | 2 |
| 1.0000E+00 | 1.0104242001E+10 | 1.01042420007E+10 | 1.010424200070E+10 | 2 |
| 2.0000E+00 | 1.0068310736E+10 | 1.00683107364E+10 | 1.006831073642E+10 | 2 |
| 3.0000E+00 | 1.0049143594E+10 | 1.00491435935E+10 | 1.004914359353E+10 | 2 |
| 4.0000E+00 | 1.0037600560E+10 | 1.00376005596E+10 | 1.003760055960E+10 | 2 |
| 5.0000E+00 | 1.0029740467E+10 | 1.00297404665E+10 | 1.002974046651E+10 | 1 |
| 7.5000E+00 | 1.0017984257E+10 | 1.00179842572E+10 | 1.001798425721E+10 | 2 |
| 1.0000E+01 | 1.0011886158E+10 | 1.00118861582E+10 | 1.001188615822E+10 | 2 |

### (b) $a = 0.01$  $B = 10^{-11}$

| | | | | |
|---|---|---|---|---|
| 0.0000E+00 | 1.0000000000E+00 | 1.00000000000E+00 | 1.000000000000E+00 | |
| 1.0000E-01 | 1.1672108377E+00 | 1.16721083771E+00 | 1.167210837715E+00 | 1 |
| 5.0000E-01 | 4.2699528441E+00 | 4.26995284405E+00 | 4.269952844053E+00 | 2 |
| 1.0000E+00 | 1.4823509385E+07 | 1.48235093847E+07 | 1.482350938467E+07 | 2 |
| 2.0000E+00 | 1.0016052907E+09 | 1.00160529072E+09 | 1.001605290724E+09 | 2 |
| 3.0000E+00 | 1.0617721285E+09 | 1.06177212854E+09 | 1.061772128536E+09 | 2 |
| 4.0000E+00 | 1.0442587702E+09 | 1.04425877016E+09 | 1.044258770157E+09 | 2 |
| 5.0000E+00 | 1.0337982900E+09 | 1.03379829001E+09 | 1.033798290011E+09 | 2 |
| 7.5000E+00 | 1.0194902855E+09 | 1.01949028546E+09 | 1.019490285462E+09 | 2 |
| 1.0000E+01 | 1.0124316714E+09 | 1.01243167142E+09 | 1.012431671425E+09 | 2 |

### (c) $a = 0.003$  $B = 10^{-11}$

| | | | | |
|---|---|---|---|---|
| 0.0000E+00 | 1.0000000000E+00 | 1.00000000000E+00 | 1.000000000000E+00 | |
| 1.0000E-01 | 1.0453716663E+00 | 1.04537166630E+00 | 1.045371666299E+00 | 1 |
| 5.0000E-01 | 1.3246619846E+00 | 1.32466198459E+00 | 1.324661984585E+00 | 2 |
| 1.0000E+00 | 2.0531949811E+00 | 2.05319498114E+00 | 2.053194981136E+00 | 1 |
| 2.0000E+00 | 2.3920639689E+01 | 2.39206396888E+01 | 2.392063968876E+01 | 2 |
| 3.0000E+00 | 8.0570978433E+08 | 8.05709784331E+08 | 8.057097843307E+08 | 1 |
| 4.0000E+00 | 3.4133407009E+08 | 3.41334070087E+08 | 3.413340700868E+08 | 1 |
| 5.0000E+00 | 3.2932662695E+08 | 3.29326626954E+08 | 3.293266269537E+08 | 2 |
| 7.5000E+00 | 3.2271921821E+08 | 3.22719218212E+08 | 3.227192182120E+08 | 1 |
| 1.0000E+01 | 3.1459364163E+08 | 3.14593641634E+08 | 3.145936416341E+08 | 1 |

### (d) $a = 0.001$  $B = 10^{-11}$

| | | | | |
|---|---|---|---|---|
| 0.0000E+00 | 1.0000000000E+00 | 1.00000000000E+00 | 1.000000000000E+00 | |
| 1.0000E-01 | 1.0147177710E+00 | 1.01471777105E+00 | 1.014717771047E+00 | 2 |
| 5.0000E-01 | 1.0898213935E+00 | 1.08982139354E+00 | 1.089821393537E+00 | 2 |
| 1.0000E+00 | 1.2136025788E+00 | 1.21360257879E+00 | 1.213602578792E+00 | 1 |
| 2.0000E+00 | 1.6144148814E+00 | 1.61441488145E+00 | 1.614414881449E+00 | 1 |
| 3.0000E+00 | 2.4471103261E+00 | 2.44711032608E+00 | 2.447110326079E+00 | 1 |
| 4.0000E+00 | 4.6439263602E+00 | 4.64392636023E+00 | 4.643926360228E+00 | 2 |
| 5.0000E+00 | 1.4163711393E+01 | 1.41637113932E+01 | 1.416371139317E+01 | |
| 7.5000E+00 | 1.2592678382E+09 | 1.25926783818E+09 | 1.259267838179E+09 | 2 |
| 1.0000E+01 | 1.5900586024E+08 | 1.59005860241E+08 | 1.590058602409E+08 | 2 |

(e)  $a = 0.1$  $B = 10^{-13}$

| | | | |
|---|---|---|---|
| 0.0000E+00 | 1.0000000000E+00 | 1.00000000000E+00 | 1.000000000000E+00 | |
| 1.0000E- 01 | 2.4733658251E+01 | 2.47336582509E+01 | 2.473365825088E+01 | 2 |
| 5.0000E- 01 | 1.5433617863E+12 | 1.54336178629E+12 | 1.543361786285E+12 | 2 |
| 1.0000E+00 | 1.0100735783E+12 | 1.01007357828E+12 | 1.010073578280E+12 | 2 |
| 2.0000E+00 | 1.0068324164E+12 | 1.00683241638E+12 | 1.006832416382E+12 | 2 |
| 3.0000E+00 | 1.0049146603E+12 | 1.00491466032E+12 | 1.004914660322E+12 | 1 |
| 4.0000E+00 | 1.0037601614E+12 | 1.00376016137E+12 | 1.003760161371E+12 | 2 |
| 5.0000E+00 | 1.0029740921E+12 | 1.00297409214E+12 | 1.002974092136E+12 | 2 |
| 7.5000E+00 | 1.0017984372E+12 | 1.00179843719E+12 | 1.001798437192E+12 | 1 |
| 1.0000E+01 | 1.0011886207E+12 | 1.00118862073E+12 | 1.001188620726E+12 | 2 |

(f)  $a = 0.01$  $B = 10^{-13}$

| | | | |
|---|---|---|---|
| 0.0000E+00 | 1.0000000000E+00 | 1.00000000000E+00 | 1.000000000000E+00 | |
| 1.0000E- 01 | 1.1672108379E+00 | 1.16721083792E+00 | 1.167210837918E+00 | 1 |
| 5.0000E- 01 | 4.2699528644E+00 | 4.26995286440E+00 | 4.269952864400E+00 | 2 |
| 1.0000E+00 | 1.4833155106E+07 | 1.48331551059E+07 | 1.483315510587E+07 | 2 |
| 2.0000E+00 | 9.2815225125E+10 | 9.28152251255E+10 | 9.281522512546E+10 | 2 |
| 3.0000E+00 | 1.0620282033E+11 | 1.06202820331E+11 | 1.062028203309E+11 | 2 |
| 4.0000E+00 | 1.0446043780E+11 | 1.04460437803E+11 | 1.044604378025E+11 | 2 |
| 5.0000E+00 | 1.0338896655E+11 | 1.03388966548E+11 | 1.033889665484E+11 | 2 |
| 7.5000E+00 | 1.0194999125E+11 | 1.01949991252E+11 | 1.019499912524E+11 | 2 |
| 1.0000E+01 | 1.0124348832E+11 | 1.01243488316E+11 | 1.012434883158E+11 | 2 |

(g)  $a = 0.003$  $B = 10^{-13}$

| | | | |
|---|---|---|---|
| 0.0000E+00 | 1.0000000000E+00 | 1.00000000000E+00 | 1.000000000000E+00 | |
| 1.0000E- 01 | 1.0453716665E+00 | 1.04537166646E+00 | 1.045371666458E+00 | 1 |
| 5.0000E- 01 | 1.3246619862E+00 | 1.32466198617E+00 | 1.324661986171E+00 | 2 |
| 1.0000E+00 | 2.0531949903E+00 | 2.05319499029E+00 | 2.053194990288E+00 | 1 |
| 2.0000E+00 | 2.3920642249E+01 | 2.39206422487E+01 | 2.392064224873E+01 | 1 |
| 3.0000E+00 | 6.3125860906E+11 | 6.31258609057E+11 | 6.312586090572E+11 | 2 |
| 4.0000E+00 | 3.3555955479E+10 | 3.35559554785E+10 | 3.355595547852E+10 | 1 |
| 5.0000E+00 | 3.2156761131E+10 | 3.21567611308E+10 | 3.215676113084E+10 | 1 |
| 7.5000E+00 | 3.2102051821E+10 | 3.21020518214E+10 | 3.210205182144E+10 | 2 |
| 1.0000E+01 | 3.1456146867E+10 | 3.14561468668E+10 | 3.145614686684E+10 | 1 |

(h)  $a = 0.001$  $B = 10^{-13}$

| | | | |
|---|---|---|---|
| 0.0000E+00 | 1.0000000000E+00 | 1.00000000000E+00 | 1.000000000000E+00 | |
| 1.0000E- 01 | 1.0147177712E+00 | 1.01471777120E+00 | 1.014717771195E+00 | 2 |
| 5.0000E- 01 | 1.0898213945E+00 | 1.08982139454E+00 | 1.089821394541E+00 | 2 |
| 1.0000E+00 | 1.2136025816E+00 | 1.21360258156E+00 | 1.213602581557E+00 | 2 |
| 2.0000E+00 | 1.6144148929E+00 | 1.61441489293E+00 | 1.614414892934E+00 | 1 |
| 3.0000E+00 | 2.4471103706E+00 | 2.44711037061E+00 | 2.447110370609E+00 | 2 |
| 4.0000E+00 | 4.6439265876E+00 | 4.64392658765E+00 | 4.643926587648E+00 | 2 |
| 5.0000E+00 | 1.4163713862E+01 | 1.41637138620E+01 | 1.416371386022E+01 | 1 |
| 7.5000E+00 | 8.3616177063E+09 | 8.36161770625E+09 | 8.361617706254E+09 | 2 |
| 1.0000E+01 | 1.6158767668E+10 | 1.61587676681E+10 | 1.615876766808E+10 | 2 |

## 6.3 Insertion Proportional to Neutron Density

A more straightforward non-linear density- dependent feedback is proportionality to the neutron density itself

$$\rho(t) = 0.1\beta N(t), \tag{18a}$$

to give the TS coefficient

$$\rho_{n,j} = 0.1\beta N_{n,j}, \quad n = 0,1,\dots. \tag{18b}$$

We choose Thermal Reactor II of Table A.IV with neutron lifetime of $5x10^{-4}$. Table 8a confirms all entries but two (off by one unit) are in perfect agreement to 12 places (for **m1** set to 2). The last two digits come into compliance when the **err** is set to $5x10^{-15}$.

This insertion is particularly interesting since the neutron density becomes unbounded in a finite time. Figure 7 shows the steep ascent with time at about $26s$. As time approaches $26.163683^+s$, the neutron density approaches infinity. Apparently, the neutron density becomes infinite in the interval

$$[26.163683588, 26.163683589],$$

which needs verification.

The sudden change in agreement between DPCATS and BEFD in the second column of Table 8b indicates that higher precision arithmetic is required to track the neutron

### Table 8a. Proportional Insertion TRII.
### ($\Lambda = 5x10^{-4}$, err = $10^{-12}$)

| $t(s)$ | BEFD/CATS(10) | BEFD/CATS(11) | BEFD/CATS(12) | conv |
|---|---|---|---|---|
| 0.0000E+00 | 1.0000000000E+00 | 1.00000000000E+00 | 1.000000000000E+00 | |
| 1.0000E-01 | 1.0817840726E+00 | 1.08178407260E+00 | 1.081784072596E+00 | 2 |
| 5.0000E-01 | 1.1435623844E+00 | 1.14356238440E+00 | 1.143562384399E+00 | 2 |
| 1.0000E+00 | 1.1676688085E+00 | 1.16766880850E+00 | 1.167668808496E+00 | 2 |
| 2.0000E+00 | 1.2079218113E+00 | 1.20792181132E+00 | 1.207921811316E+00 | 2 |
| 5.0000E+00 | 1.3199424643E+00 | 1.31994246426E+00 | 1.319942464258E+00 | 1 |
| 1.0000E+01 | 1.5296094953E+00 | 1.52960949528E+00 | 1.529609495280E+00 | 2 |
| 1.5000E+01 | 1.8179563592E+00 | 1.81795635915E+00 | 1.817956359151E+00 | 1 |
| 2.0000E+01 | 2.3139472813E+00 | 2.31394728130E+00 | 2.313947281298E+00 | 2 |
| 2.5000E+01 | 4.0904249107E+00 | 4.09042491066E+00 | 4.090424910661E+00 | 2 |
| 2.6000E+01 | 9.1631391167E+00 | 9.16313911667E+00 | 9.163139116673E+00 | 1 |

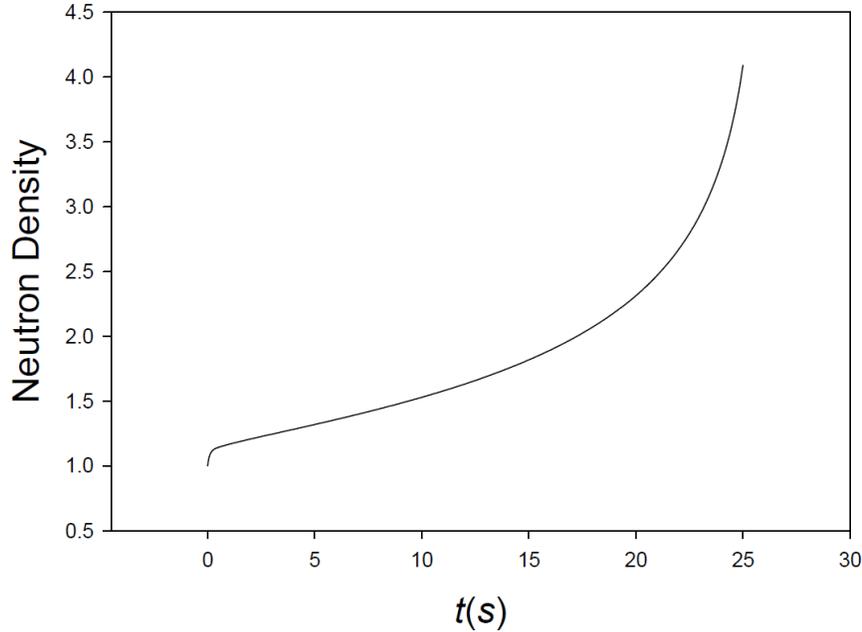

Fig. 7. Neutron density approach to infinity.

**Table 8b. Asymptotic approach of neutron density to infinity**
**($\Lambda = 2x10^{-5}$, err = $8x10^{-15}$)**

| $t(s)$ | DPCATS(10) | QPCATS(10) | BEFD(10) |
|---|---|---|---|
| 2.6000000E+01 | 9.1631391167E+00 | 9.1631391167E+00 | 9.1631391167E+00 |
| 2.6100000E+01 | 1.6795369146E+01 | 1.6795369146E+01 | 1.6795369146E+01 |
| 2.6160000E+01 | 2.1373243263E+02 | 2.1373243263E+02 | 2.1373243263E+02 |
| 2.6163000E+01 | 1.1299299378E+03 | 1.1299299378E+03 | 1.1299299378E+03 |
| 2.6163600E+01 | 9.2047441**902**E+03 | 9.2047441929E+03 | 9.2047441929E+03 |
| 2.6163630E+01 | 1.4354947**695**E+04 | 1.4354947701E+04 | 1.4354947701E+04 |
| 2.6163683E+01 | 1.3064063**145**E+06 | 1.3064063682E+06 | 1.3064063682E+06 |

density as it approaches infinity. This is evident when QPCATS (third application), is included in column three and matches BEFD to ten places. When we evaluate at time closer to the singularity as shown in Table 8c, QPCATS responds with larger and larger numbers, but BEFD fails to converge. We conclude that CATS can do this exceptional case but only with QP arithmetic.

**Table 8c: Approach to singularity.**

| $t(s)$ | QPCATS(10) |
|---|---|
| 26.163683000000 | 1.3064063682E+06 |
| 26.163683588000 | 1.2101591483E+09 |
| 26.163683588568 | 1.1401156760E+10 |

## 7. ADDITIONAL PRESCRIBED REACTIVITY INSERTIONS

During the past fifty years, few if any new reactivity insertions have appeared in the literature as benchmarks. That is about to change as we investigate several new prescribed reactivities as suggested by the CATS algorithm and confirmed by the BEFD algorithm. Since CATS requires reactivity in the form of a power series (TS) to be input into the recurrence of Eqs(4), virtually any common function is a suitable reactivity candidate making the possibilities practically limitless. In essence, only the $n^{\text{th}}$ TS coefficient $\rho_{n,j}$ of the reactivity is necessary. In this regard, we review several ways of generating TS. Since BEFD only requires the reactivity as a given function, we can simultaneous confirm the CATS benchmark; and, in so doing establish a suite of new PKE benchmarks. Aside from including Doppler, the reactivities considered do not necessarily represent true transients and are to be viewed simply as benchmarks.

Our approach is to choose the following four example reactivities inserted into TRIII:

| Insertion | $\rho(t)$ | $\rho_0$ | $\alpha$ | $\mu$ | $\sigma$ | $\gamma$ |
|---|---|---|---|---|---|---|
| Exponential | $\rho_0 e^{-\alpha t}$ | \$1 | 0.25,0.5,1 | --- | --- | --- |
| Sine | $\rho_0 \sin(\gamma t)$ | \$1 | --- | --- | --- | 0.5,1,2 |
| Shifted Gaussian | $\dfrac{\rho_0}{\sqrt{2\pi}\sigma} e^{-(t-\mu)^2/2\sigma^2}$ | \$1 | --- | 0.1,1,3,10 10 | 1 0.5,1,2,3 | --- |
| Bessel function | $\rho_0 J_0(\alpha t)$ | \$1 \$0.2,0.4, 0.6,0.8 | 0.5,1,5 1 | --- | --- | --- |

and determine their TS coefficients using each of four methods: differentiation, McLaurin Series, recurrence and from the literature. From these examples, we then construct additional reactivities in the following section.

### 7.1 Selected reactivity insertions into TRIII
### a. Exponential insertion: By Differentiation

Consider an exponential insertion

$$\rho(t) = \rho_0 e^{-\alpha t}, \ t_{j-1} \le t \le t_j . \tag{19}$$

The most straightforward way to find the Taylor coefficients for reactivity is simply differentiate $n$ times to give the TS coefficient at $t_{j-1}$

$$\rho_{n,j-1} = \rho_0 \frac{(-1)^n}{n!} \alpha^n e^{-\alpha t_{j-1}}, \ \ n = 0,1,\dots . \tag{20}$$

Incorporating the coefficient into the CATS algorithm through the recurrence of Eqs(4) and by comparing the resulting neutron density to that of BEFD for exponential reactivity insertion of $\rho_0 = \$1$ and $\alpha = 1$ into TRIII, Table 9a1 indicates perfect agreement.

### Table 9a1. Exponential Insertion into TRIII
($\rho_0 = \$1$, $\Lambda = 2x10^{-5}$, $\alpha = 1$)

| t(s) | BEFD/CATS(10) | BEFD/CATS(11) | BEFD/CATS(12) | conv |
|---|---|---|---|---|
| 0.0000E+00 | 1.0000000000E+00 | 1.00000000000E+00 | 1.000000000000E+00 | |
| 1.0000E -02 | 4.4510555338E+00 | 4.45105553376E+00 | 4.451055533761E+00 | 1 |
| 1.0000E -01 | 1.6774521759E+01 | 1.67745217587E+01 | 1.677452175867E+01 | 1 |
| 1.0000E+01 | 1.8023302385E+00 | 1.80233023845E+00 | 1.802330238451E+00 | 2 |
| 5.0000E+01 | 1.5361360685E+00 | 1.53613606854E+00 | 1.536136068539E+00 | 1 |
| 1.0000E+02 | 1.5119197781E+00 | 1.51191977813E+00 | 1.511919778127E+00 | 2 |

The insertion, a step from critical followed by an exponential falloff, is shown in Figs. 8a,b for increasing falloff. Note the insertion is not a prompt jump because of delay caused by the delayed neutrons at early time as shown in Fig. 8b.

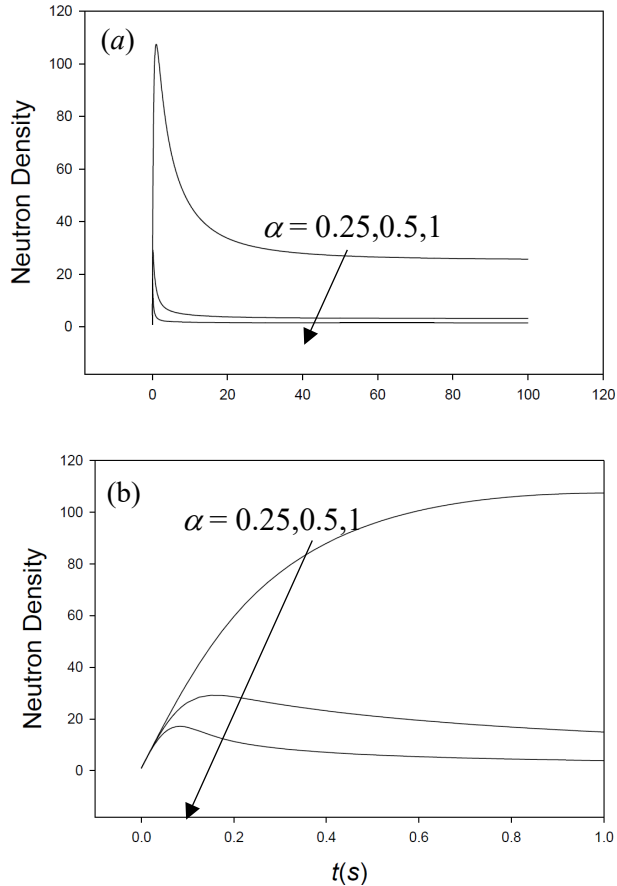

Fig. 8. a. Exponential insertion ($\alpha$ =0.25, 0.5 1).
b. Exponential insertion just after critical.

To extend the exponential insertion to a more complicated one, consider

$$\rho(t) = \rho_0 e^{\alpha \sin(\gamma t)}, \; t_{j-1} \le t \le t_j, \tag{21a}$$

with derivative

$$\frac{d\rho(t)}{dt} = \rho_0 \alpha \gamma \cos(\gamma t) \rho(t). \tag{21b}$$

For the $n^{th}$ derivative, perform consecutive differentiation and apply Leibnitz's formula for the $n^{th}$ derivative of a product [30] (to play a central role here)

$$\frac{d^n \rho(t)}{dt^n} = \rho_0 \alpha \gamma \frac{d^{n-1} \rho(t)}{dt^{n-1}} \frac{d}{dt} \cos(\gamma t) \rho(t) \tag{22a}$$

$$\rho^{(n)}(t) = \rho_0 \alpha \gamma \sum_{l=0}^{n-1} \frac{(n-1)!}{l!(n-1-l)!} \frac{d^{n-1-l} \cos(\gamma t)}{dt^{n-1-l}} \rho^{(l)}(t). \tag{22b}$$

Since

$$\frac{d^{n-1-l} \cos(\gamma t)}{dt^{n-1-l}} = \operatorname{Re}\left[\frac{d^{n-1-l} e^{i\gamma t}}{dt^{n-1-l}}\right] = \operatorname{Re}\left[(i\gamma)^{n-1-l} e^{i\gamma t}\right], \tag{23a}$$

and taking the real part gives

$$\rho_{n,j-1} = \frac{\rho_0 \alpha \gamma}{n} \sum_{l=0}^{n-1} \frac{\gamma^{n-1-l}}{(n-1-l)!} \cos\left(\gamma t_{j-1} + \pi(n-1-l)/2\right) \rho_{l,j-1}. \tag{23b}$$

For $\alpha = -1$ and $\gamma = 0.08$, Table 9a2 shows perfect agreement with BEFD. Figures 9a,b show the rather odd behavior of the neutron density response. Apparently, the exponential controls the early time and the sine the later time. Eventually. each curve will head toward infinity.

**Table 9a2. Exponential Insertion with sine argument into TRIII.**
**($\rho_0 = \$1$, $\Lambda = 2x10^{-5}$, $\alpha = -1$, $\gamma = 0.8$)**

| $t(s)$ | BEFD/CATS(10) | BEFD/CATS(11) | BEFD/CATS(12) | conv |
|---|---|---|---|---|
| 0.0000E+00 | 1.0000000000E+00 | 1.00000000000E+00 | 1.000000000000E+00 | 1 |
| 1.0000E-02 | 4.4625138585E+00 | 4.46251385845E+00 | 4.462513858452E+00 | 1 |
| 1.0000E-01 | 1.9923186789E+01 | 1.99231867894E+01 | 1.992318678935E+01 | 1 |
| 5.0000E-01 | 9.0340927784E+00 | 9.03409277845E+00 | 9.034092778445E+00 | 1 |
| 1.0000E+00 | 6.2109993534E+00 | 6.21099935338E+00 | 6.210999353383E+00 | 1 |
| 2.0000E+00 | 5.1729513267E+00 | 5.17295132667E+00 | 5.172951326669E+00 | 1 |
| 3.0000E+00 | 7.3922904966E+00 | 7.39229049662E+00 | 7.392290496620E+00 | 1 |
| 4.0000E+00 | 1.3161689705E+03 | 1.31616897046E+03 | 1.316168970458E+03 | 1 |

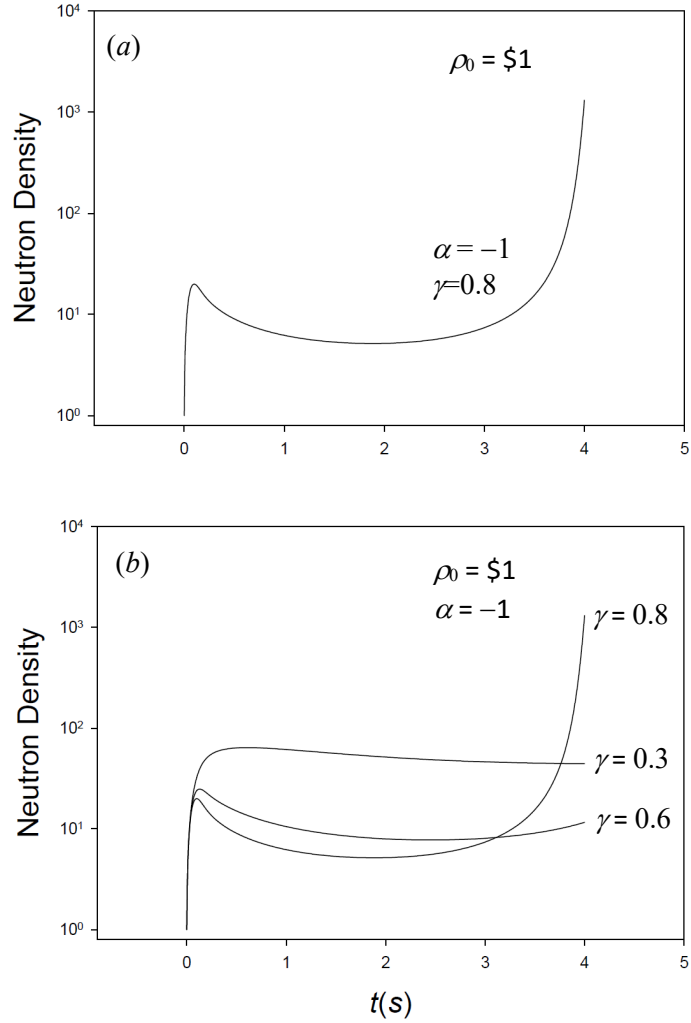

Fig. 9. Neutron density for exponential sine argument variation in ($a$) $\alpha$ and ($b$) $\gamma$.

## b. Sine Insertion: By McLaurin Series

We considered the sine insertion previously, but here we apply an alternative method to find the $n^{th}$ derivative using the McLaurin series.

For reactivity

$$\rho(t) = \rho_0 \sin(\gamma t), \ t_{j-1} \leq t \leq t_j,$$  (24a)

one can write

$$\rho(t) = \rho_0 \operatorname{Im} e^{i\gamma t}.$$  (24b)

Then, from the McLaurin series for the exponential,

$$e^{i\gamma t} = \sum_{n=0}^{\infty} \frac{(i\gamma)^n}{n!} \, t^n, \tag{25a}$$

and shifting $t$ by $t_{j-1}$ gives

$$e^{i\gamma(t-t_{j-1})} = e^{i\gamma t} e^{-i\gamma t_{j-1}} = \sum_{n=0}^{\infty} \frac{(i\gamma)^n}{n!} \left(t - t_{j-1}\right)^n. \tag{25b}$$

Then, a second expression for the exponential is

$$e^{i\gamma t} = \sum_{n=0}^{\infty} \gamma^n \frac{i^n e^{i\gamma t_{j-1}}}{n!} \left(t - t_{j-1}\right)^n \tag{26a}$$

and taking the imaginary part

$$\sin\left(\gamma t\right) = \sum_{n=0}^{\infty} \frac{\gamma^n}{n!} \operatorname{Im}\left(i^n e^{i\gamma t_{j-1}}\right) \left(t - t_{j-1}\right)^n, \tag{26b}$$

which is the TS for sine in the interval $t_{j-1} \leq t \leq t_j$. Since the coefficient of $\left(t - t_{j-1}\right)^n$ is the TS coefficient

$$\frac{1}{n!} \frac{d^n}{dt^n} \sin\left(\gamma t\right) \bigg|_{t_{j-1}} = \frac{\gamma^n}{n!} \operatorname{Im}\left(i^n e^{i\gamma t_{j-1}}\right) \tag{27a}$$

and

$$\rho_{n,j-1} = \rho_0 \frac{\gamma^n}{n!} \sin\left(\gamma t_{j-1} + \pi n / 2\right). \tag{27b}$$

For $\rho_0 = \$1$, $\gamma = 0.5$, we find perfect agreement between the two benchmarks for sine insertion as recorded in Table 9b1. In addition, Fig. 10a shows the expected induced oscillation in the neutron density for several values of $\gamma$. Also, note the increasing density and large values at large time. The increase comes about because of the cycling positive and negative reactivity over time.

| $t(s)$ | BEFD/CATS(10) | BEFD/CATS(11) | BEFD/CATS(12) | conv |
|---|---|---|---|---|
| 0.0000E+00 | 1.0000000000E+00 | 1.00000000000E+00 | 1.000000000000E+00 | |
| 1.0000E -02 | 1.0036278921E+00 | 1.00362789209E+00 | 1.003627892088E+00 | 1 |
| 1.0000E -01 | 1.0519635883E+00 | 1.05196358833E+00 | 1.051963588330E+00 | 2 |
| 1.0000E+01 | 2.4486238003E+06 | 2.44862380028E+06 | 2.448623800278E+06 | 2 |
| 5.0000E+01 | 1.7938198780E+26 | 1.79381987799E+26 | 1.793819877990E+26 | 2 |
| 7.5000E+01 | 2.4033270507E+39 | 2.40332705068E+39 | 2.403327050683E+39 | 2 |
| 1.0000E+02 | 3.2374576338E+52 | 3.23745763381E+52 | 3.237457633808E+52 | 2 |

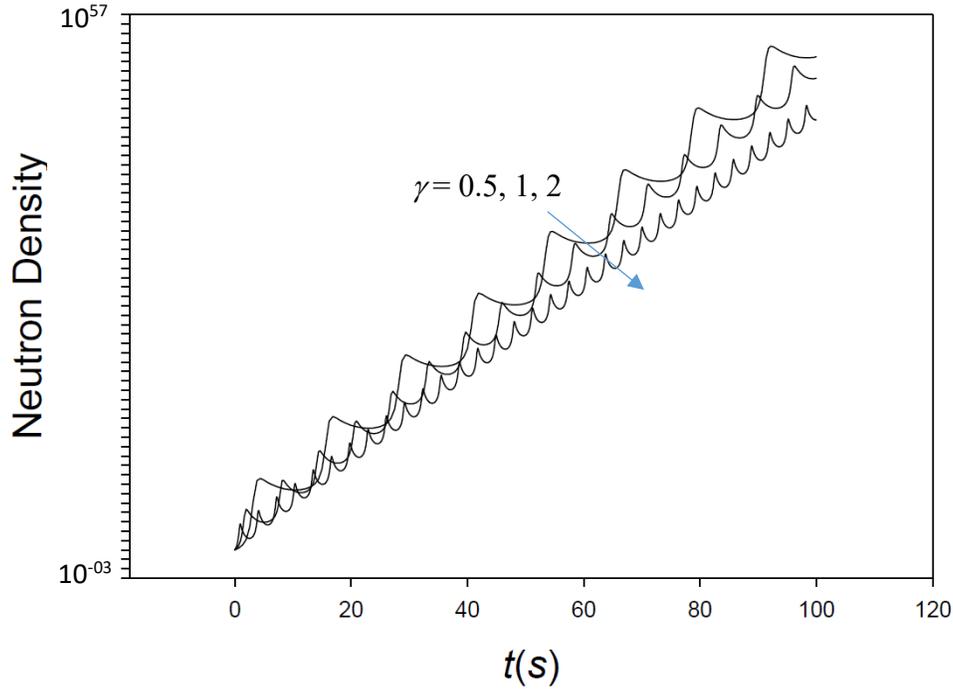

Fig. 10a. Sine insertion for $\gamma$ = 0.5, 1, 2.

As a variation, now consider the following more complicated waveform for reactivity:

$$\rho(t) = \rho_0 \left[ g_1 \sin(\gamma_1 t) + g_2 \cos(\gamma_2 t) \right], \ t_{j-1} \leq t \leq t_j. \tag{28}$$

Taking the real part of Eq(26a) gives

$$\cos(\gamma t) = \sum_{n=0}^{\infty} \frac{\gamma^n}{n!} \text{Re}\left( i^n e^{i\gamma t_{j-1}} \right) \left( t - t_{j-1} \right)^n \tag{29a}$$

to form the TS coefficient from Eqs (26b) and (29a)

$$\rho_{n,j-1} = \frac{\rho_0}{n!} \left[ g_1 \gamma_1^n \, \mathrm{Im}\left( i^n e^{i\gamma_1 t_{j-1}} \right) + g_2 \gamma_2^n \, \mathrm{Re}\left( i^n e^{i\gamma_2 t_{j-1}} \right) \right]. \tag{29b}$$

Assuming

$$
\begin{aligned}
\rho_0 &= \$1 \\
g_1 &= 0.75 \\
g_2 &= 0.25 \\
\gamma_1 &= 0.5 \\
\gamma_2 &= 0.5/\pi,
\end{aligned}
\tag{30}
$$

on comparison with BEFD, perfect agreement results as shown in Table 9b2.

**Table 9b2. More complicated Sine Insertion into TRIII.**
**($\rho_0 = \$1$, $\Lambda = 2\mathrm{x}10^{-5}$, $g_1 = 0.75$, $g_2 = 0.25$, $\gamma_1 = 0.5$, $\gamma_2 = 0.5/\pi$)**

| $t(s)$ | BEFD/CATS(10) | BEFD/CATS(11) | BEFD/CATS(11) | conv |
|---|---|---|---|---|
| 0.0000E+00 | 1.0000000000E+00 | 1.00000000000E+00 | 1.000000000000E+00 | |
| 1.0000E-02 | 1.3140584016E+00 | 1.31405840163E+00 | 1.314058401630E+00 | 2 |
| 1.0000E-01 | 1.4201345306E+00 | 1.42013453061E+00 | 1.420134530610E+00 | 2 |
| 1.0000E+01 | 6.5775879848E+04 | 6.57758798483E+04 | 6.577587984827E+04 | 2 |
| 5.0000E+01 | 2.1229415327E+12 | 2.12294153270E+12 | 2.122941532702E+12 | 1 |
| 7.5000E+01 | 4.8634888388E+12 | 4.86348883880E+12 | 4.863488838801E+12 | 2 |
| 1.0000E+02 | 1.8785888606E+20 | 1.87858886056E+20 | 1.878588860558E+20 | 2 |

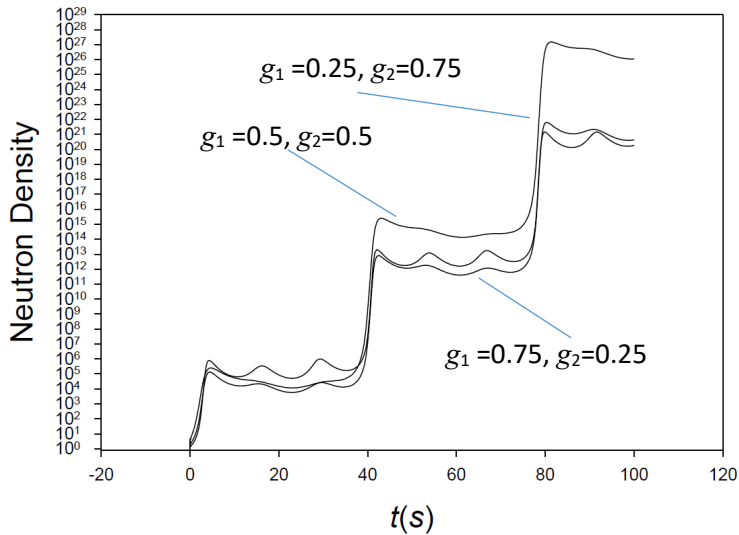

Fig. 10b. Neutron density for more complicated waveform.

Figure 10b shows the more complicated neutron density response for the waveform of Eq(28) as two superimposed waves. This opens up the possibility of Fourier series representations of reactivity with a wide range of function possibilities.

### c. Shifted Gaussian Insertion: By Recurrence

The shifted Gaussian insertion for mean $\mu$ and dispersion $\sigma$

$$\rho(t) = \frac{\rho_0}{\sqrt{2\pi}\sigma} e^{-(t-\mu)^2/2\sigma^2} \tag{31}$$

presents a similar challenge to the insertion of Eq(21a). Only by splitting the $n^{th}$ derivative into two consecutive derivatives, one of first order

$$\rho^{(n)}(t) = \frac{\rho_0}{\sqrt{2\pi}\sigma} \frac{d^{n-1}}{dt^{n-1}} \frac{d}{dt} e^{-(t-\mu)^2/2\sigma^2}, \tag{32a}$$

with the first order derivative performed explicitly, one finds

$$\rho^{(n)}(t) = -\frac{1}{\sigma^2} \frac{d^{n-1}}{dt^{n-1}} \left[ (t-\mu) \frac{\rho_0}{\sqrt{2\pi}\sigma} e^{-(t-\mu)^2/2\sigma^2} \right]. \tag{32b}$$

Then, identifying the Gaussian reactivity in the term in brackets

$$\rho^{(n)}(t) = -\frac{1}{\sigma^2} \frac{d^{n-1}}{dt^{n-1}} \left[ (t-\mu)\rho(t) \right] \tag{33a}$$

and from Leibnitz's product differentiation

$$\rho^{(n)}(t) = -\frac{1}{\sigma^2} \sum_{l=0}^{n-1} \frac{(n-1)!}{l!(n-1-l)!} \left[ \frac{d^l}{dt^l}(t-\mu) \right] \rho^{(n-1-l)}(t), \tag{33b}$$

there results the following recurrence at $t = t_{j-1}$:

$$\rho_{j-1}^{(n)} \equiv \rho_{n,j-1} = -\frac{1}{\sigma^2} \left[ (t_{j-1}-\mu)\rho_{j-1}^{(n-1)} + (n-1)\rho_{j-1}^{(n-2)} \right] \tag{33c}$$

or the TS coefficient by recurrence is

$$\rho_{n,j-1} = -\frac{1}{n\sigma^2} \left[ (t_{j-1}-\mu)\rho_{n-1,j-1} + \rho_{n-2,j-1} \right] \tag{33d}$$

initiated by

$$\rho_{0,j} = \frac{\rho_0}{\sqrt{2\pi}\,\sigma}\, e^{-\left(t_j - \mu\right)^2 / 2\sigma^2}. \tag{33e}$$

Table 9c shows the comparison for a shifted Gaussian reactivity centered about $\mu = 3$ with dispersion of unity and $\rho_0 = \$1$. We see perfect agreement to 12 places for the nominal case in Table 9c.

### Table 9c. Shifted Gaussian Insertion for TRIII
### ($\rho_0 = \$1$, $\Lambda = 2x10^{-5}$, $\mu = 3$, $\sigma = 1$)

| $t(s)$ | BEFD/CATS(10) | BEFD/CATS(11) | BEFD/CATS(12) | $conv$ |
|---|---|---|---|---|
| 0.0000E+00 | 1.0000000000E+00 | 1.00000000000E+00 | 1.000000000000E+00 | |
| 1.0000E-02 | 1.0044221872E+00 | 1.00442218720E+00 | 1.004422187197E+00 | 2 |
| 1.0000E-01 | 1.0061403508E+00 | 1.00614035081E+00 | 1.006140350815E+00 | 2 |
| 1.0000E+00 | 1.0646324990E+00 | 1.06463249900E+00 | 1.064632498998E+00 | 2 |
| 1.0000E+01 | 1.2421861201E+00 | 1.24218612010E+00 | 1.242186120104E+00 | 2 |
| 2.0000E+01 | 1.1814969337E+00 | 1.18149693366E+00 | 1.181496933662E+00 | 2 |
| 2.5000E+01 | 1.1690820343E+00 | 1.16908203429E+00 | 1.169082034289E+00 | 2 |

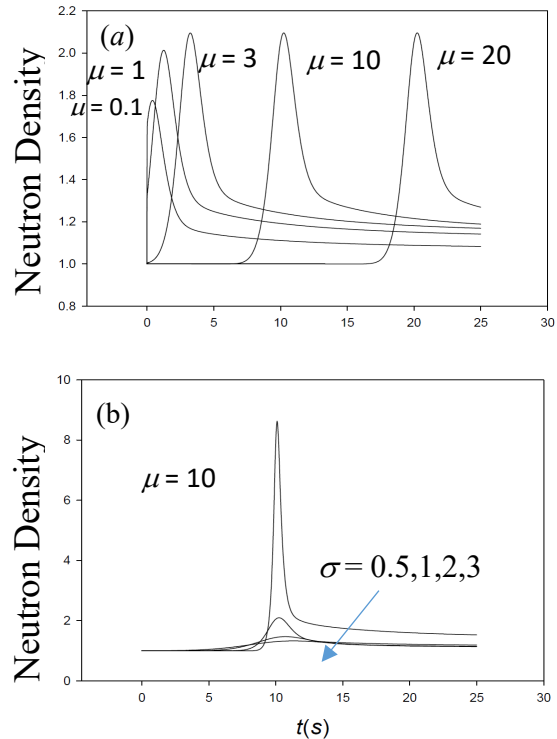

Fig. 11. (a) Gaussian insertion for $\mu = 0.1,1,\ 3,10,20$ and $\sigma = 1$ in TRIII.
(b) Dispersion reduction for $\sigma = 0.5,1,2,3$ at $\mu = 10$.

Figure 11a shows a pulse train advancing in time centered from 0.1 to 10$s$ with $\rho_0 = \$1$. Apparently, the pulse reaches a fix shape at sufficiently long time. In contrast,

Fig. 11b shows the dispersion of a fixed pulse centered at $t = 10s$ with $\sigma$ ranging from 0.5 to 3 forcing the pulse height to increase with decreasing dispersion since the Gaussian is constant on integration.

### d. Bessel function insertion: From the Literature

For the last example, we consider a zeroth order Bessel function insertion

$$\rho(t) = \rho_0 J_0(\alpha t). \tag{34}$$

For this unusual insertion whose $n^{th}$ derivative for $m = 0$ comes from the literature [31]

$$\rho_j^{(n)} = \frac{d^n J_m(\alpha t)}{dt^n} = \rho_0 \left[\frac{\alpha}{2}\right]^n \sum_{k=0}^{n} (-1)^k \binom{n}{k} J_{2k-n+m}(\alpha t) \tag{35a}$$

giving the TS coefficient when $m = 0$

$$\rho_{n,j} = \frac{1}{n!} \frac{d^n J_0(\alpha t)}{dt^n}\bigg|_{t_j} = \frac{\rho_0}{n!} \left[\frac{\alpha}{2}\right]^n \sum_{k=0}^{n} (-1)^k \binom{n}{k} J_{2k-n}(\alpha t_j). \tag{35b}$$

On comparison with BEFD, we see only one discrepancy in the twelfth place for $\alpha = 1$ and $\rho_0 = \$0.2$ shown in Table 9d.

### Table 9d. Bessel function Insertion for TRIII
### ($\rho_0 = \$0.2$, $\Lambda = 2x10^{-5}$, $\alpha = 1$)

| $t(s)$ | BEFD/CATS(10) | BEFD/CATS(11) | BEFD/CATS(12) | conv |
|---|---|---|---|---|
| 0.0000E+00 | 1.0000000000E+00 | 1.00000000000E+00 | 1.000000000000E+00 | |
| 1.0000E- 02 | 1.2353216419E+00 | 1.23532164189E+00 | 1.235321641890E+00 | 2 |
| 1.0000E- 01 | 1.2613107936E+00 | 1.26131079362E+00 | 1.261310793617E+00 | 1 |
| 1.0000E+00 | 1.2648939690E+00 | 1.26489396900E+00 | 1.264893969004E+00 | 2 |
| 1.0000E+01 | 9.8636673361E -01 | 9.86366733608E -01 | 9.863667336081E -01 | 1 |
| 1.0000E+02 | 1.0261817333E+00 | 1.02618173332E+00 | 1.026181733323E+00 | 1 |

Fig. 12a shows the increase in the frequency of oscillation in time with decreasing $\alpha$ and for $\$1$ insertion. The oscillations come from the functional behavior of the Bessel function. In Fig. 12b, we see a growing first pulse followed by smaller pulses increasing in amplitude with reactivity. Note any order Bessel function can be a benchmark.

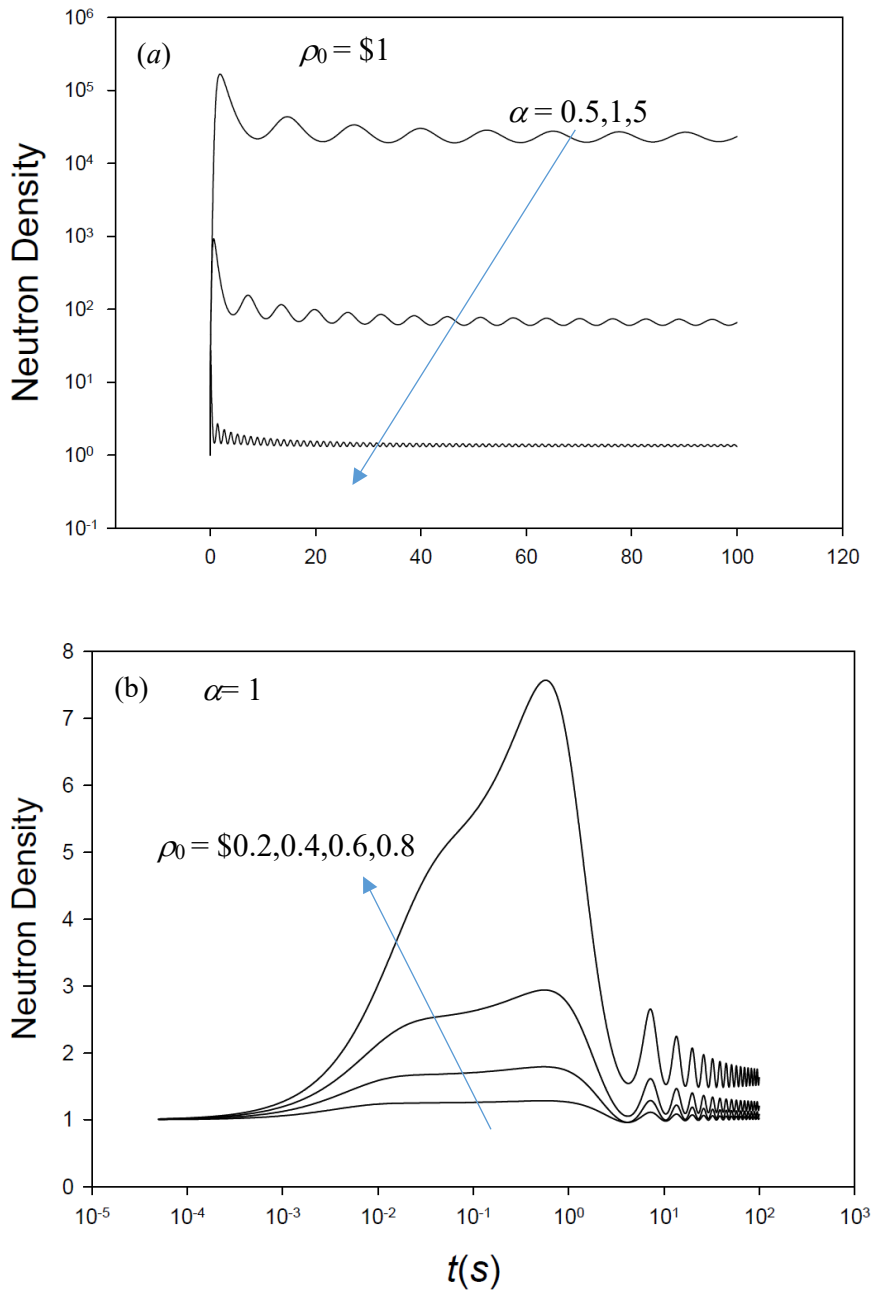

Fig. 12. (a) Bessel functions oscillations in time for $\rho_0 = \$1$, $\alpha = 0.5, 1, 5$.
(b) Increase in density response with reactivity for $\rho_0 = \$0.2, 0.4, 0.6, 0.8$ with $\alpha = 1$.

## 7.2 Combination reactivity insertions into TRIII

The following set of insertions take the form

$$\rho(t) = \rho_0 f(t) g(t), \tag{36a}$$

which from Leibnitz's differentiation most conveniently gives the derivative

$$\rho^{(n)}(t) = \rho_0 \sum_{l=0}^{n} \frac{n!}{l!(n-l)!} f^{(n-l)}(t) g^{(l)}(t). \qquad (36b)$$

Therefore, the TS coefficient of reactivity for the combination becomes at $t_j$

$$\rho_{n,j} = \rho_0 \sum_{l=0}^{n} f_{n-l,j} g_{l,j} \qquad . \qquad (37)$$

to be introduced into Eqs(4).

## a. Oscillating/Exponential Insertion
The oscillating exponentially damped insertion is

$$\rho(t) = \rho_0 \sin(\gamma t) e^{-\alpha t} \qquad (38a)$$

for which

$$f(t) = \sin(\gamma t), \ g(t) = e^{-\alpha t}. \qquad (38b)$$

However, before applying Eq(38a) directly in Eq(37), the following alternative as applied previously simplifies the derivation of the TS coefficient.

Since the insertion is equivalent to

$$\rho(t) = \rho_0 \operatorname{Im} e^{i\gamma t} e^{-\alpha t}, \qquad (39a)$$

combination of the exponentials give

$$\rho(t) = \rho_0 \operatorname{Im} e^{i\gamma t - \alpha t} \qquad (39b)$$

and on differentiation

$$\rho^{(n)}(t) = \rho_0 \operatorname{Im}\left( (-1)^n (\alpha - i\gamma)^n e^{-(\alpha - i\gamma)t} \right). \qquad (39c)$$

With simplification, the TS coefficient at $t_j$ becomes

$$\rho_{n,j} = \rho_0 \frac{(-1)^n}{n!} \left(\alpha^2 + \gamma^2\right)^{n/2} \sin\left(\gamma t_j - n \tan^{-1}\left(\gamma / \alpha\right)\right) e^{-\alpha t_j}. \qquad (40)$$

Again, we find perfect agreement as shown in Table 10a and observe the oscillations of Fig. 12, which shows the influence of the exponential factor of the insertion to suppress the increase caused by the sine factor.

**Table 10a. Oscillating/Exponential Insertion in TRIII**
**($\rho_0 = \$1, \Lambda = 2x10^{-5}, \alpha = 0.1, \gamma = 2$)**

| $t(s)$ | BEFD/CATS(10) | BEFD/CATS(11) | BEFD/CATS(12) | $conv$ |
|---|---|---|---|---|
| 0.0000E+00 | 1.0000000000E+00 | 1.00000000000E+00 | 1.000000000000E+00 | |
| 1.0000E-02 | 1.0146190215E+00 | 1.01461902149E+00 | 1.014619021486E+00 | 2 |
| 1.0000E-01 | 1.2396514745E+00 | 1.23965147446E+00 | 1.239651474460E+00 | 2 |
| 1.0000E+01 | 6.1311855874E+00 | 6.13118558741E+00 | 6.131185587406E+00 | 2 |
| 5.0000E+01 | 3.0355853322E+00 | 3.03558533219E+00 | 3.035585332191E+00 | 2 |
| 1.0000E+02 | 2.9453018081E+00 | 2.94530180809E+00 | 2.945301808091E+00 | 2 |

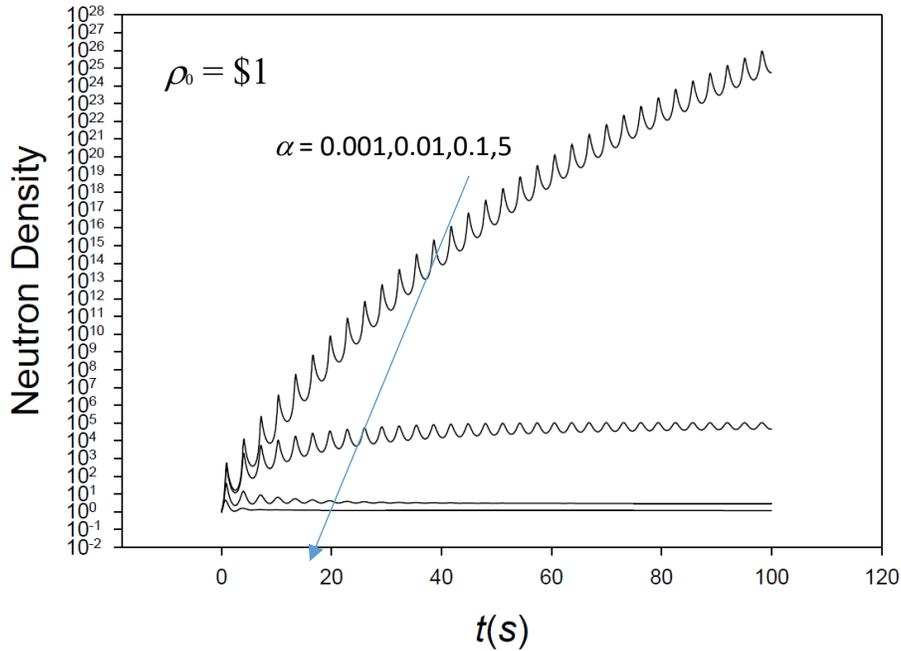

Fig. 13. Oscillating/Exponential Insertion in TRIII.

Now back to the original form of Eq(36a) with application of Leibnitz's differentiation, we find from Eq(37)

$$f_{n,j} = \frac{\gamma^n}{n!} \mathrm{Im}\left(i^n e^{i\gamma t_j}\right) \qquad (41a)$$

$$g_{n,j} = \frac{(-1)^n}{n!} \alpha^n e^{-\alpha t_j}, \quad n = 0,1,\dots. \qquad (41b)$$

and therefore

$$\rho_{n,j} = \rho_0 \sum_{l=0}^{n} f_{n-l,j} g_{l,j},$$  (42)

which gives identical results (not shown) as Table 10a.

## b. Oscillating/Shifted- Gaussian /Exponential Insertion
A more complicated reactivity involves sine, Gaussian and exponential

$$\rho(t) = \rho_0 f(t) g(t) h(t),$$  (43a)

where

$$f(t) = \sin(\gamma t), \ g(t) = e^{-\gamma t}, \ h(t) = \frac{e^{-(t-\mu)^2/2\sigma^2}}{\sqrt{2\pi}\sigma}.$$  (43b)

From the previous example, one application of Leibnitz's differentiation gives the TS coefficient for $f(t)g(t)$ as

$$s_{n,j} \equiv \sum_{l=0}^{n} f_{n-l,j} g_{l,j},$$  (44a)

where

$$f_{n,j} = \frac{\gamma^n}{n!} \text{Im}\left(i^n e^{i\gamma t_j}\right)$$  (44b)

$$g_{n,j} = \frac{(-1)^n}{n!} \alpha^n e^{-\alpha t_j}.$$  (44c)

A second application of Leibnitz's differentiation then gives the triple product reactivity associated with Eqs(43a,b) as

$$\rho_{n,j} \equiv \rho_0 \sum_{l=0}^{n} s_{n-l,j} h_{l,j},$$  (45)

where $h_{n,j}$ is the recurrence for the Shifted- Gaussian distribution found in §7.1c. Table 10b confirms all places of CATS relative to BEFD. For a Gaussian pulse centered at $5s$ with a spread of $1s$, Fig. 13 shows the imprinted oscillations over the pulse width.

**Table 10b. Oscillating/Gaussian/Exponential Insertion for TRIII**
**($\rho_0 = \$1$, $\Lambda = 2x10^{-5}$, $\alpha = 0.01$, $\sigma = 1.0$, $\mu = 5$, $\gamma = 20$)**

| $t(s)$ | BEFD/CATS(10) | BEFD/CATS(11) | BEFD/CATS(12) | conv |
|---|---|---|---|---|
| 0.0000E+00 | 1.0000000000E+00 | 1.00000000000E+00 | 1.000000000000E+00 | |
| 1.0000E-02 | 1.0000002230E+00 | 1.00000022302E+00 | 1.000000223022E+00 | 1 |
| 1.0000E-01 | 1.0000022892E+00 | 1.00000228916E+00 | 1.000002289161E+00 | 1 |
| 5.0000E+00 | 8.4047931778E-01 | 8.40479317785E-01 | 8.404793177847E-01 | 2 |
| 1.0000E+01 | 1.0215505490E+00 | 1.02155054897E+00 | 1.021550548969E+00 | 2 |
| 1.5000E+01 | 1.0173288001E+00 | 1.01732880011E+00 | 1.017328800110E+00 | 2 |

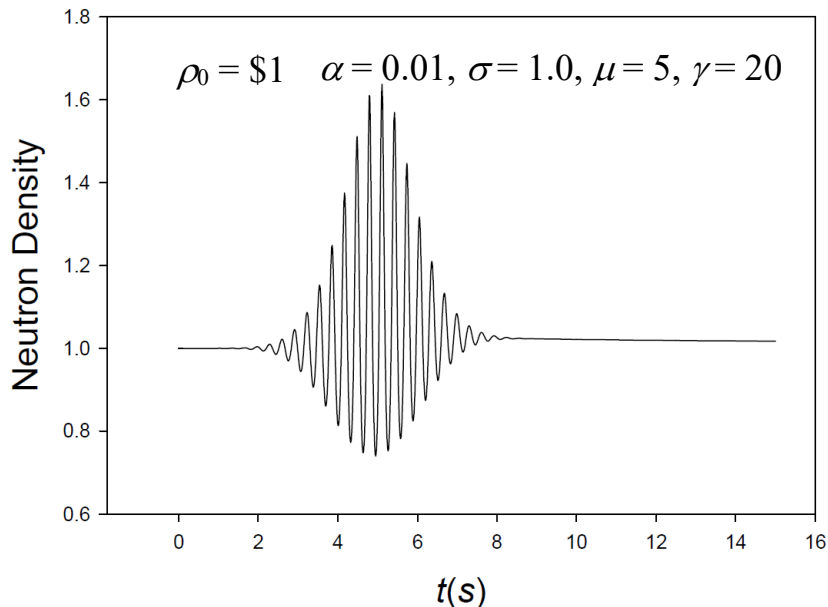

Fig. 14. Imprinted oscillations over the pulse width.

## 8. VARIATIONS ON DOPPLER FEEDBACK

In this section, we consider several generalizations with the Doppler effect. The generalizations admit TS for reactivity that conform to the CATS algorithm. The variations are similar to § 6.1 and 6.2 but more sophisticated and therefore one should not expect the 10- or 11- place standard to hold as readily. For this reason, we seek only 9- or 10- place benchmarks, which still qualify as extreme benchmarks comparable to or with more precision than generally found in the literature. Several 11- place benchmarks will also appear. All examples will be for Thermal Reactor IV with kinetic parameters from Table A.V in Appendix A.

### 8.1. Doppler Feedback: Internal Factor

We first consider a modified Doppler feedback with an internal factor in the Doppler term

$$\rho(t) = \rho_0 - B \int\limits_0^t dt' f(t') N(t').$$ (46)

One proceeds as before by finding the $n^{th}$ derivative of the reactivity according to

$$\rho^{(n)}(t) = \frac{d^{n-1}}{dt^{n-1}} \frac{d\rho(t)}{dt} = -B \frac{d^{n-1}}{dt^{n-1}} f(t) N(t)$$ (47a)

to give

$$\rho^{(n)}(t) = \rho_0 \delta_{n,0} - B \sum_{l=0}^{n-1} \frac{(n-1)!}{l!(n-1-l)!} f^{(n-1-l)}(t) N^{(l)}(t).$$ (47b)

The TS coefficient is therefore

$$\rho_{n,j-1} = \rho_0 \delta_{n,0} - \frac{B}{n} \sum_{l=0}^{n-1} f_{n-1-l,j-1} N_{l,j-1}.$$ (48)

As a demonstration, we consider several factors with step reactivities.

**a.** Exponential Factor
If

$$f(t) \equiv e^{-\alpha t},$$ (49a)

then

$$f_n = \frac{(-1)^n}{n!} \alpha^n e^{-\alpha t},$$ (49b)

and the TS coefficient becomes

$$\rho_{n,j-1} = \rho_0 \delta_{n,0} - \frac{B}{n} e^{-\alpha t_{j-1}} \sum_{l=0}^{n-1} (-1)^{(n-1-l)} \frac{\alpha^{(n-1-l)}}{(n-1-l)!} N_{l,j-1}$$ (49c)

with Doppler exponent $\alpha = 0.1$. Table 11a1 shows no discrepancies for 9 or 10-places for the NIPL with *l2* set to 12 to give a greater possibility for convergence within the partitions. In the eleventh place, two entries do not conform. When *l2* is set to 15, only the second discrepancy remains. Wrangling did not work in this case, so we accept the 10-place benchmark.

Figure 15 shows how $\alpha$ enables control of the Doppler effect to shut down the step transient. For $\alpha > 0$, Doppler dies away relatively quickly and offers minimal transient reduction. For $\alpha < 0$, the Doppler term is larger and when sufficiently large provides enough negative reactivity to turn the transient around. To the author's knowledge, showing control of the Doppler in this fashion is new.

**Table 11a1. Doppler Feedback with an Internal Exponential Factor**
**($\rho_0 = 0.1$, $\Lambda = 5 \times 10^{-5}$, $\alpha = 0.1$)**

| $t(s)$ | BEFD/CATS(9) | BEFD/CATS(10) | BEFD/CATS(11) |
|---|---|---|---|
| 0.0000E+00 | 1.000000000E+00 | 1.0000000000E+00 | 1.00000000000E+00 |
| 1.0000E+00 | 1.146784429E+00 | 1.1467844285E+00 | 1.14678442851E+00 |
| 1.0000E+01 | 1.332167947E+00 | 1.3321679468E+00 | 1.33216794681E+00 |
| 2.0000E+01 | 1.500952047E+00 | 1.5009520468E+00 | 1.50095204684E+00 |
| 3.0000E+01 | 1.668934194E+00 | 1.6689341938E+00 | 1.66893419381E+00 |
| 4.0000E+01 | 1.846248311E+00 | 1.8462483107E+00 | 1.84624831075E+00 |
| 5.0000E+01 | 2.037493519E+00 | 2.0374935191E+00 | 2.03749351915E+00 |
| 6.0000E+01 | 2.245722110E+00 | 2.2457221099E+00 | 2.24572210988E+00 |
| 7.0000E+01 | 2.473470810E+00 | 2.4734708104E+00 | 2.47347081040E+00 |
| 8.0000E+01 | 2.723138348E+00 | 2.7231383483E+00 | 2.72313834826E+00 |
| 9.0000E+01 | 2.997163664E+00 | 2.9971636640E+00 | 2.99716366400E+00 |
| 1.0000E+02 | 3.298124366E+00 | 3.2981243655E+00 | 3.29812436554E+00 |

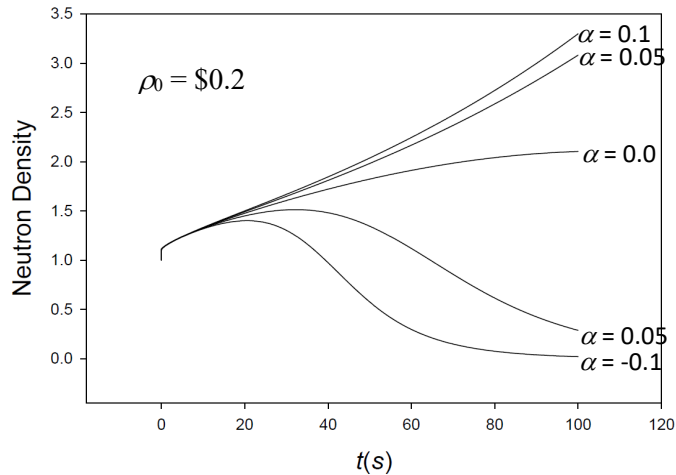

Fig. 15. Doppler control through $\alpha$.

**b.** Sine factor
For

$$f\left(t\right) \equiv \sin\left(\gamma t\right),\qquad\qquad(50a)$$

then

$$f_n(t) = \sin(\gamma t + \pi n / 2) \qquad (50b)$$

and the TS coefficient with Doppler is

$$\rho_{n,j-1} = \rho_0 \delta_{n,0} - \frac{B}{n} \sum_{l=0}^{n-1} \frac{\gamma^{(n-1-l)}}{(n-1-l)!} \sin(\gamma t + \pi(n-1-l)/2) N_{l,j-1}. \qquad (50c)$$

Table 11a2 indicates that NIPL for **err** of $10^{-12}$ and **l2** at 12 has no discrepancies for 9 and 10 places, but four occur for 11 places. Most notably, when **err** is set to $10^{-13}$, all 11 places now agree.

**Table 11a2. Doppler Feedback with an Internal Sine Factor**
**($\rho_0 = 0.2$, $\Lambda = 5x10^{-5}$, $\gamma = 5$, $B = 2.5x10^{-6}$)**

| $t(s)$ | BEFD/CATS(9) | BEFD/CATS(10) | BEFD/CATS(11) |
|---|---|---|---|
| 0.0000E+00 | 1.000000000E+00 | 1.0000000000E+00 | 1.00000000000E+00 |
| 1.0000E+00 | 1.343645243E+00 | 1.3436452430E+00 | 1.34364524297E+00 |
| 1.0000E+01 | 1.936737853E+00 | 1.9367378533E+00 | 1.93673785328E+00 |
| 2.0000E+01 | 2.647546946E+00 | 2.6475469456E+00 | 2.64754694562E+00 |
| 4.0000E+01 | 4.667892768E+00 | 4.6678927677E+00 | 4.66789276772E+00 |
| 6.0000E+01 | 8.076561475E+00 | 8.0765614747E+00 | 8.07656147474E+00 |
| 8.0000E+01 | 1.391093956E+01 | 1.3910939558E+01 | 1.39109395584E+01 |
| 1.0000E+02 | 2.391730138E+01 | 2.3917301378E+01 | 2.39173013782E+01 |

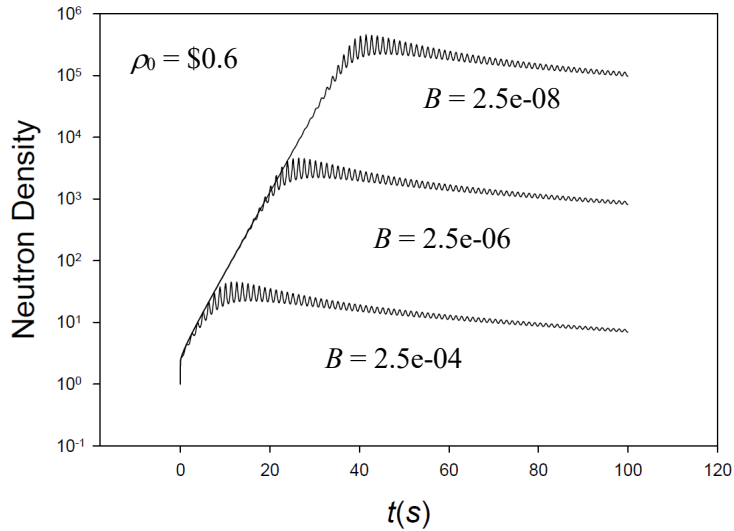

Fig. 16. Doppler shut down by sine internal factor.

Figure 16 shows how the $0.6 reactivity insertion shuts down with a sine factor of frequency 3 coupled to a decreasing Doppler coefficient. The power oscillation is a unique feature of the sine reactivity trace.

## 8.2. Doppler Feedback: External Factor

As another variation, consider the factor external to the Doppler term

$$\rho(t) = \rho_0 - Bf(t)\int_0^t dt' N(t').$$ (51)

Separating the integral into past and current Doppler contributions

$$\rho(t) = \rho_0 - Bf(t)\left[\int_0^{t_{j-1}} dt' N(t') + \int_{t_{j-1}}^t dt' N(t')\right];$$ (52a)

and noting when $t = t_{j-1}$

$$-B\int_0^{t_{j-1}} dt' N(t') = -\frac{\rho_0 - \rho(t_{j-1})}{f(t_{j-1})}$$ (52b)

gives

$$\rho(t) = \rho_0 - \frac{f(t)}{f(t_{j-1})}\left[\rho_0 - \rho(t_{j-1})\right] - Bf(t)\int_{t_{j-1}}^t dt' N(t').$$ (52c)

On differentiating $n$ times by application of Leibnitz's formula

$$\rho^{(n)}(t) = \rho_0\delta_{n,0} - \frac{f^{(n)}(t)}{f(t_{j-1})}\left[\rho_0 - \rho(t_{j-1})\right] - B\sum_{l=0}^n \frac{n!}{l!(n-l)!}f^{(n-l)}(t)\frac{d^l}{dt^l}\int_{t_{j-1}}^t dt' N(t'),$$ (53a)

separating out the first term of the sum and from Eq(14b), there results
$$\rho^{(n)}(t) =$$
$$= \rho_0\delta_{n,0} - \frac{f^{(n)}(t)}{f(t_{j-1})}\left[\rho_0 - \rho(t_{j-1})\right] - B\sum_{l=1}^n \frac{n!}{l!(n-l)!}f^{(n-l)}(t)N^{(l-1)}(t) - Bf^{(n)}(t)\int_{t_{j-1}}^t dt' N(t').$$ (53b)

As a consequence of continuity

$$\rho(t_{j-1}) \equiv \rho_{0,j-1};$$ (54a)

and at $t = t_{j-1}$

$$\rho_{n,j-1} = \rho_0 \delta_{n,0} - \frac{f_{n,j-1}}{f_{0,j-1}} \Big[ \rho_0 - \rho_{0,j-1} \Big] - B \sum_{l=1}^{n} \frac{1}{l} f_{n-l,j-1} N_{l-1,j-1}. \qquad (54b)$$

Now for a demonstration of several external factors.

**a.** External Exponential Factor
If

$$f(t) \equiv e^{-\alpha t}, \qquad (55a)$$

then

$$f^{(n)}(t) = \frac{(-1)^n}{n!} \alpha^n e^{-\alpha t} \qquad (55b)$$

and the TS coefficient is

$$\rho_{n,j-1} = \rho_0 \delta_{n,0} - \frac{(-1)^{n-1}}{n!} \alpha^n e^{-\alpha t_{j-1}} \Big[ \rho_0 - \rho_{0,j-1} \Big] - B e^{-\alpha t_{j-1}} \sum_{l=1}^{n} (-1)^{(n-l)} \frac{\alpha^{(n-l)}}{l(n-l)!} N_{l-1,j-1}. $$

$$(55c)$$

**Table 11b1. Doppler Feedback with an External Exponential Factor**
**($\rho_0 = 0.1$, $\Lambda = 5 \times 10^{-5}$, $\alpha = 1.0$, $B = 2.5 \times 10^{-6}$)**

| $t(s)$ | BEFD/CATS(9) | BEFD/CATS(10) | BEFD/CATS(11) |
|---|---|---|---|
| 0.0000E+00 | 1.000000000E+00 | 1.0000000000E+00 | 1.00000000000E+00 |
| 1.0000E+00 | 1.147148287E+00 | 1.1471482871E+00 | 1.14714828712E+00 |
| 1.0000E+01 | 1.341891344E+00 | 1.3418913444E+00 | 1.34189134439E+00 |
| 2.0000E+01 | 1.524321242E+00 | 1.5243212420E+00 | 1.52432124196E+00 |
| 4.0000E+01 | 1.903306872E+00 | 1.9033068715E+00 | 1.90330687151E+00 |
| 6.0000E+01 | 2.347276140E+00 | 2.3472761398E+00 | 2.34727613982E+00 |
| 8.0000E+01 | 2.884599866E+00 | 2.8845998660E+00 | 2.88459986600E+00 |
| 1.0000E+02 | 3.540257001E+00 | 3.5402570007E+00 | 3.54025700071E+00 |

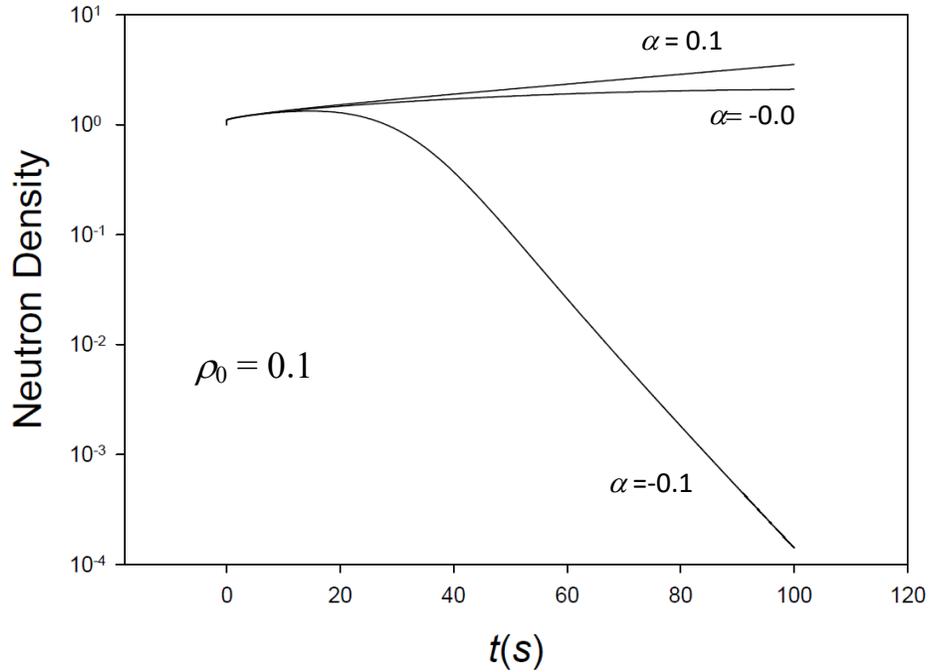

Fig. 17. Doppler control through $\alpha$ for external exponential factor.

Table 11b1 indicates four discrepancies in the 11th place for **err** set to $10^{-11}$; but when set to $3 \times 10^{-12}$, all discrepancies disappear. Figure 17 shows the control of the Doppler.

**b.** External Gaussian Factor
For

$$f\left(t\right) \equiv e^{-\alpha t^2},$$

(56a)

then

$$f^{(n)}\left(t\right) = -2\alpha \frac{d^{n-1}}{dt^{n-1}} t e^{-\alpha t^2}$$

(56b)

which gives the recurrence

$$f_{n,j-1} = -\frac{2\alpha}{n}\left(t_{j-1} f_{n-1,j-1} + f_{n-2,j-1}\right).$$

(56c)

initiated by

$$f_{0.j-1} = -2\alpha e^{-\alpha t_{j-1}^2}.$$ (56d)

The TS coefficient with Doppler is Eq.(54b).

All digits to 11 places agree as shown in Table 11b2 for NIPL with *l2* set to 12 and **m1** set to 1.

**Table 11b2. Doppler Feedback with an External Gaussian Factor**
($\rho_0 = 0.1$, $\Lambda = 5x10^{-5}$, $\alpha = 0.1$, $B = 2.5x10^{-6}$)

| *t(s)* | BEFD/CATS(9) | BEFD/CATS(10) | BEFD/CATS(11) |
|---|---|---|---|
| 0.0000E+00 | 1.000000000E+00 | 1.0000000000E+00 | 1.00000000000E+00 |
| 1.0000E+00 | 1.146811100E+00 | 1.1468111005E+00 | 1.14681110050E+00 |
| 1.0000E+01 | 1.341459632E+00 | 1.3414596325E+00 | 1.34145963249E+00 |
| 2.0000E+01 | 1.523933344E+00 | 1.5239333435E+00 | 1.52393334353E+00 |
| 4.0000E+01 | 1.902881669E+00 | 1.9028816688E+00 | 1.90288166882E+00 |
| 6.0000E+01 | 2.346768735E+00 | 2.3467687351E+00 | 2.34676873510E+00 |
| 8.0000E+01 | 2.883983064E+00 | 2.8839830641E+00 | 2.88398306409E+00 |
| 1.0000E+02 | 3.539503427E+00 | 3.5395034268E+00 | 3.53950342681E+00 |

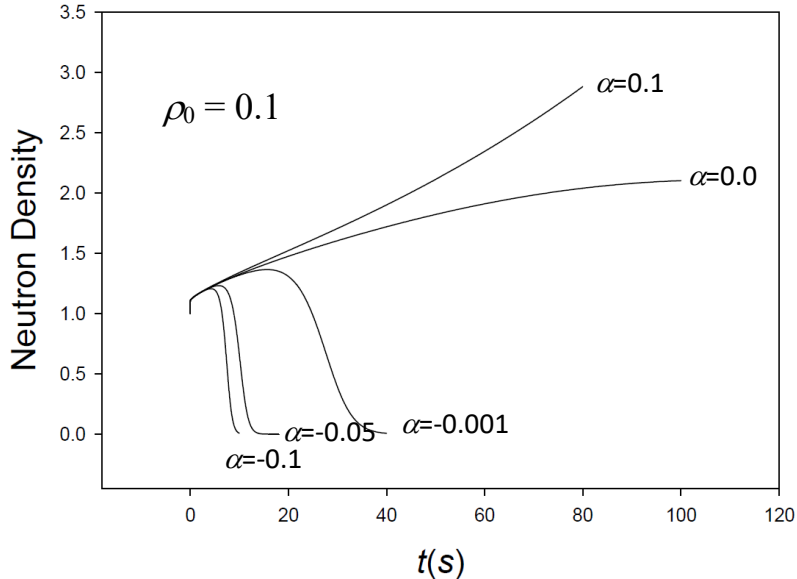

Fig. 18. Doppler shut down for Gaussian factor.

Figure 18 shows a much faster shut down of the transient for the Gaussian factor than for the exponential.

**c.** External Exponential/Sine Factor
For

$$f\left(t\right) \equiv e^{\alpha \sin(\gamma t)};$$ (57a)

then from Eq(22b), with $\rho_0$ set to unity

$$f_{0,j-1} = e^{\alpha \sin(\gamma t_{j-1})}$$

$$f_{n,j-1} = \frac{\alpha \gamma}{n} \sum_{l=0}^{n-1} \frac{\gamma^{n-1-l}}{(n-1-l)!} \cos\left(\gamma t_{j-1} + \pi(n-1-l)/2\right) f_{l,j-1}, \qquad (57b)$$

and the TS coefficient with Doppler follows from Eq.(54b).

As shown in Table 11b3, agreement to 10 places is achieved for the choices of $\rho_0$, $\alpha$, $\gamma$ given with **err** = $10^{-13}$ and **m1** = 0. One entry in the $11^{th}$ place does not conform and resists wrangling. Figure 19 shows how the Doppler oscillations limit the transient by waves of ever decreasing amplitude according to $\alpha$. Along with case in §8.1a, this case is one of the most difficult as we do not find full eleven-place agreement since one entry is off by one unit. In addition, the time of computation is relatively long. While 9 and 10 places are relatively stable with respect to the choice of $\alpha$ and $\gamma$, 11 places are not.

**Table 11b3. Doppler Feedback with an External Exponential/Sine Factor**
**($\rho_0$ = 0.5, $\Lambda$ = 5x10$^{-5}$, $\alpha$ = 0.1, $\gamma$ = 1.0, $B$ = 2.5x10$^{-6}$)**

| $t(s)$ | BEFD/CATS(9) | BEFD/CATS(10) | BEFD/CATS(11) |
|---|---|---|---|
| 0.0000E+00 | 1.000000000E+00 | 1.0000000000E+00 | 1.00000000000E+00 |
| 1.0000E+00 | 2.667056408E+00 | 2.6670564081E+00 | 2.66705640812E+00 |
| 1.0000E+01 | 1.314604787E+01 | 1.3146047865E+01 | 1.31460478652E+01 |
| 2.0000E+01 | 3.618881147E+01 | 3.6188811468E+01 | 3.61888114684E+01 |
| 4.0000E+01 | 3.522080112E+01 | 3.5220801122E+01 | 3.52208011221E+01 |
| 6.0000E+01 | 2.001492629E+01 | 2.0014926290E+01 | 2.00149262898E+01 |
| 8.0000E+01 | 1.230882458E+01 | 1.2308824580E+01 | 1.23088245795E+01 |
| 1.0000E+02 | 7.648715601E+00 | 7.6487156009E+00 | 7.64871560087E+00 |

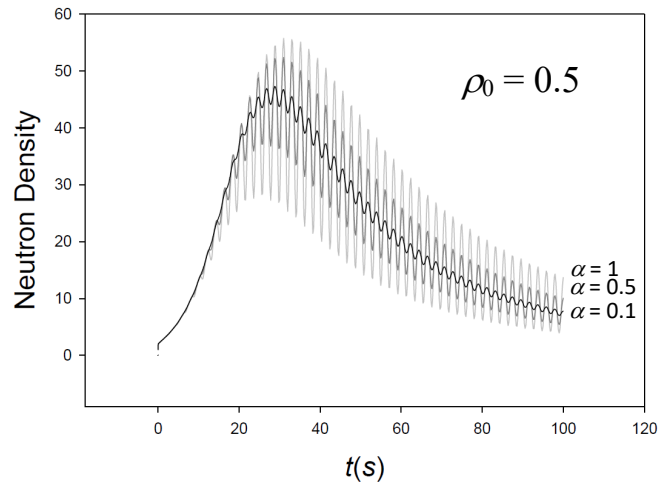

Fig. 19. Decreasing wave amplitude with $\alpha$.

## 8.3. Non-linear Doppler Exponential Convolution

Finally, we consider a non-linear convolution form of reactivity insertion more general than found in either §6.1 and §6.2 and the last two insertions

$$\rho(t) = \rho_0 - B\int_0^t dt' e^{-\alpha(t-t')} N(t').$$ (58)

Isolating the reactivity up to and during the current time interval gives

$$\rho(t) = \rho_0 - B\int_0^{t_{j-1}} dt' e^{-\alpha(t-t')} N(t') - B\int_{t_{j-1}}^t dt' e^{-\alpha(t-t')} N(t'),$$ (59a)

which when differentiated becomes

$$\frac{d\rho(t)}{dt} = -B\left[ -\alpha\int_0^t dt' e^{-\alpha(t-t')} N(t') + N(t) - \alpha\int_{t_{j-1}}^t dt' e^{-\alpha(t-t')} N(t') \right].$$ (59b)

From Eq(59a)

$$B\int_0^{t_{j-1}} dt' e^{-\alpha(t-t')} N(t') = \rho_0 - \rho(t) - B\int_{t_{j-1}}^t dt' e^{-\alpha(t-t')} N(t'),$$ (60a)

introduced into Eq(59b), we find

$$\frac{d\rho(t)}{dt} = \alpha\left[ \rho_0 - \rho(t) - B\int_{t_{j-1}}^t dt' e^{-\alpha(t-t')} N(t') + B\int_{t_{j-1}}^t dt' e^{-\alpha(t-t')} N(t') \right] - BN(t)$$ (60b)

and with cancellation

$$\frac{d\rho(t)}{dt} = -\alpha\rho(t) - BN(t) + \alpha\rho_0.$$ (60c)

The $n^{th}$ derivative is therefore

$$\frac{d^{n-1}}{dt^{n-1}}\frac{d\rho(t)}{dt} = -\alpha\rho^{(n-1)}(t) - BN^{(n-1)}(t) \qquad (61a)$$

giving the TS coefficient at $t_{j-1}$ as the simple recurrence

$$\rho_{n,j-1} = -\frac{1}{n}\Big[\alpha\rho_{n-1,j-1} + BN_{n-1,j-1}\Big]. \qquad (61b)$$

### Table 11c. Non-linear Exponential Convolution
#### ($\rho_0$ = \$0.5, $\Lambda$ = 5x10⁻⁵, $\alpha$ = −0.1)

| $t(s)$ | BEFD/CATS(9) | BEFD/CATS(10) | BEFD/CATS(11) |
|---|---|---|---|
| 0.0000E+00 | 1.000000000E+00 | 1.0000000000E+00 | 1.00000000000E+00 |
| 1.0000E+00 | 2.667280469E+00 | 2.6672804688E+00 | 2.66728046880E+00 |
| 1.0000E+01 | 1.246922241E+01 | 1.2469224213E+01 | 1.24692224129E+01 |
| 2.0000E+01 | 2.288076579E+01 | 2.2880765790E+01 | 2.28807657905E+01 |
| 3.0000E+01 | 1.084068838E+01 | 1.0840688376E+01 | 1.08406883761E+01 |
| 4.0000E+01 | 2.380413310E+00 | 2.3804133104E+00 | 2.38041331042E+00 |
| 5.0000E+01 | 4.619709053E-01 | 4.6197090527E-01 | 4.61970905269E-01 |
| 6.0000E+01 | 1.021326019E-01 | 1.0213260191E-01 | 1.02132601909E-01 |
| 7.0000E+01 | 2.525905634E-02 | 2.5259056339E-02 | 2.52590563387E-02 |
| 8.0000E+01 | 6.654819628E-03 | 6.6548196281E-03 | 6.65481962814E-03 |
| 9.0000E+01 | 1.813583428E-03 | 1.8135834280E-03 | 1.81358342804E-03 |
| 1.0000E+02 | 5.036670171E-04 | 5.0366701706E-04 | 5.03667017064E-04 |

For a step reactivity of \$0.5 and $\alpha$ = −0.1, Table 11c indicates complete agreement to 11 digits. The results came from the NIPL for **m1** = 2 and **n0**, the number of elements in the W-e window, set to 15. The same settings allow all but three entries for a 12-place benchmark (not shown).

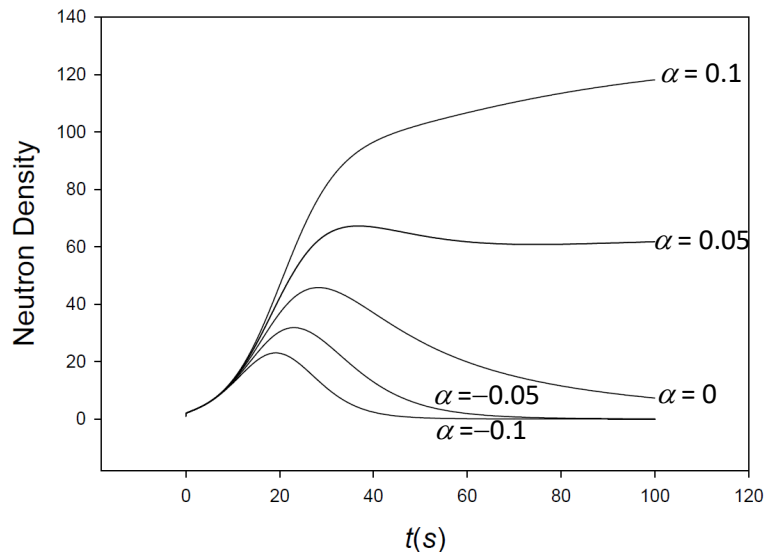

Fig. 20. Control of Doppler through the convolution insertion.

Figure 20 shows the behavior of the Doppler by the convolution, which is similar to the previous feedbacks with internal and external factors in §7.1a,b and 8.2a.

**CONCLUSION**

This presentation is a tale of two benchmarks- one(CATS) a Taylor series and the other (BEFD) a finite difference. We considered extreme benchmarking including continuous analytical continuation, partitioning, adaptivity, convergence acceleration, numerical "wrangling" and benchmark confirmation to demonstrate that DPCATS is an extreme benchmark able to resolve the PKEs to at least 10-places or better. The challenge was met by application of the high precision (to 14 or more places) independent BEFD benchmark based on a backward Euler's finite difference. With the precision of BEFD firmly established, BEFD thereafter became the standard to which we compare the CATS algorithm. Table 12 lists a summary of the 24 extreme benchmarks generated by BEFD during this study. A table of 10, 11 or 12-places accompanies each benchmark for reference. The second column refers to the number of places in agreement between the two algorithms. The number in parenthesis for some entries is the number of entries that did not comply with the indicated precision. For example, 12(2) indicates 12 place precision for all but two entries. To not overstate the precision, only the precision primarily from minor wrangling is quoted. Admittedly, some guesswork was necessary to arrive at the final digit counts, but the guesses were based on common sense sensitivity, which is expected to be a part of any benchmark study. The last column gives the time estimates of computation, which may not be the most efficient, but certainly are competitive in comparison to the methods found in the literature. All calculations were performed on a 2.6/i7 GHz Dell Precision laptop.

**Table 12: Summary of Benchmarks Presented.**

| Reactivity Insertion | CATS Precision | CPU($s$) |
|---|---|---|
| 1a.    $0.5 Step reactivity | 12 | 6.2 |
| 2a,b.  Negative Step:–$0.5/–$1 Step reactivities | 12 | 0.08 |
| 3a.    Ramp Reactivity: $0.1/$s$ ramp | 12 | 0.02 |
| 3b.    Ramp Reactivity: $1/$s$ ramp | 12(1) | 0.21 |
| 4.     Zig- Zag Insertion | 12 | 0.02 |
| 5a.    1 delayed group sinusoidal reactivity insertion | 12(2) | 107 |
| 5b.    6 delayed group sinusoidal reactivity insertion | 12(1) | 4.5 |
| 6.     Step insertion with Doppler | 12 | 3.8 |
| 7.     Keepin's Compensated Ramp Insertion | 12 | 0.67 |
| 8a.    Proportional Insertion | 12 | 0.02 |
| 9a1.   Exponential insertion: By Differentiation | 12 | 0.23 |
| 9a2.   Exponential Insertion with sine argument | 12 | 0.03 |
| 9b1.   Sine Insertion: McLaurin series | 12 | 1.4 |
| 9b2.   More complicated Sine Insertion | 12 | 1.25 |
| 9c.    Shifted Gaussian Insertion: By Recurrence | 12 | 0.04 |
| 9d.    Bessel function insertion: From the Literature | 12(1) | 5.2 |
| 10a.   Oscillating/Exponential Insertion | 11 | 1.1 |
| 10b.   Oscillating/Shifted- Gaussian/Exponential Insertion | 11 | 0.13 |
| 11a1.  Doppler Feedback with an Internal Exponential Factor | 11(1) | 5.9 |
| 11a2.  Doppler Feedback with an Internal Sine Factor | 11 | 25.6 |
| 11b1.  Doppler Feedback with an External Exponential Factor | 11 | 6.9 |
| 11b2.  Doppler Feedback with an External Gaussian Factor | 11 | 1.5 |
| 11b3.  Doppler Feedback with an External Exponential/Sine Factor | 11(1) | 46 |
| 11c.   Non-linear Exponential Convolution | 11 | 3.8 |

In the author's experience, there has never been a PKE comparison like the one presented here. One only need to ask: When have this many PKE benchmarks ever before been published? Confidence in the BEFD benchmark is what makes the difference. Some may argue that BEFD may not be as precise as claimed and therefore the study could be flawed. Certainly, BEFD imprecision cannot be ruled out for every benchmark, but referring to the high precision comparisons of BEFD cited in the literature for a variety of PKE algorithms and cases, it is not hard to argue otherwise. Our study shows the CATS algorithm for all but two cases (22 benchmarks) to be a fully 11- place benchmark; and 16 of those, 5 entries from being 12-place benchmarks. Thus, CATS indeed qualifies as an extreme benchmark.

# APPENDIX A: Reactor Kinetic Parameters

## Table A.I. Fast Reactor I (FRI)

$\Lambda = 10^{-7}s$

| $i$ | $\beta_i$ | $\lambda_i$ |
|---|---|---|
| 1 | 0.0001672 | 0.0129 |
| 2 | 0.001232 | 0.0311 |
| 3 | 0.0009504 | 0.134 |
| 4 | 0.001443 | 0.331 |
| 5 | 0.0004534 | 1.26 |
| 6 | 0.000154 | 3.21 |
| | $\beta = 0.0044$ | |

## Table A.II. Thermal Reactor I (TRI)

$\Lambda = 5x10^{-4}s$

| $i$ | $\beta_i$ | $\lambda_i$ |
|---|---|---|
| 1 | 0.000285 | 0.0127 |
| 2 | 0.0015975 | 0.0317 |
| 3 | 0.00141 | 0.115 |
| 4 | 0.0030525 | 0.311 |
| 5 | 0.00096 | 1.40 |
| 6 | 0.000195 | 3.87 |
| | $\beta = 0.00750$ | |

## Table A.III. Thermal Reactor II (TRII)

$\Lambda = 5x10^{-4}s$

| $i$ | $\beta_i$ | $\lambda_i$ |
|---|---|---|
| 1 | 0.000215 | 0.0124 |
| 2 | 0.001424 | 0.0305 |
| 3 | 0.001274 | 0.111 |
| 4 | 0.002568 | 0.301 |
| 5 | 0.000748 | 1.14 |
| 6 | 0.000273 | 3.01 |
| | $\beta = 0.006502$ | |

**Table A.IV. Thermal Reactor III (TRIII)**

| $\Lambda = 2x10^{-5}s$ | | |
|:---:|:---:|:---:|
| $i$ | $\beta_i$ | $\lambda_i$ |
| 1 | 0.000266 | 0.0127 |
| 2 | 0.001491 | 0.0317 |
| 3 | 0.001316 | 0.115 |
| 4 | 0.002849 | 0.311 |
| 5 | 0.000896 | 1.40 |
| 6 | 0.000182 | 3.87 |
| $\beta = 0.00700$ | | |

**Table A.V. Thermal Reactor IV (TRIV)**

| $\Lambda = 5x10^{-5}s$ | | |
|:---:|:---:|:---:|
| $i$ | $\beta_i$ | $\lambda_i$ |
| 1 | 0.00021 | 0.0124 |
| 2 | 0.00141 | 0.0305 |
| 3 | 0.00127 | 0.111 |
| 4 | 0.00255 | 0.301 |
| 5 | 0.00074 | 1.14 |
| 6 | 0.00027 | 3.00 |
| $\beta = 0.000645$ | | |

# Appendix B Input Parameter Lists by Case

## Table B .Input parameter lists all benchmark comparisons

| Table # | err | ne | ne1 | *l1* | *l2* | K | m1 | *CPU(s)* (DPCATS/ /QPBEFD) |
|---------|-----|-----|------|------|------|-----|------|------------------------------|
| **1a** | $10^{-24}$ | 1 | 1 | 4 | 11 | 15 | 0 | 232/36 |
| **1b** | $10^{-12}$ | 1 | 1 | 4 | 11 | 15 | 0 | 6.2/2.7 |
| **1c** | $10^{-13}$ | 1 | 1 | 4 | 11 | 15 | 3 | 11.3/5 |
| **1d** | $10^{-12}$ | 1 | 1 | 4 | 11 | 15 | 0 | 6.4/3 |
| **2a,b** | $10^{-12}$ | 1 | 1 | 4 | 11 | 15 | 2 | 0.08/2.2 |
| **3a** | $10^{-12}$ | 1 | 1 | 4 | 11 | 15 | 0 | 0.02/20 |
| **3b** | $8x10^{-15}$ | 16 | 16 | 2 | 11 | 15 | 2 | 0.75/160 |
| **4** | $10^{-12}$ | 1 | 1 | 4 | 11 | 15 | 2 | 0.02/7.6 |
| **5a** | $10^{-12}$ | 1 | 1 | 4 | 11 | 15 | 3 | 107/0.83 |
| **5b** | $10^{-12}$ | 1 | 1 | 5 | 11 | 15 | 1 | 4.5/366 |
| **6a** | $10^{-14}$ | 1 | 1 | 4 | 11 | 15 | 1 | 3.4/29 |
| **6b** | $10^{-12}$ | 1 | 1 | 4 | 12 | 15 | 2 | 0.17/33 |
| **6c** | $10^{-12}$ | 1 | 1 | 4 | 11 | 15 | 2 | 0.22/39 |
| **7a-h** | $10^{-12}/5x10^{-13}$ | 1 | 1 | 4 | 11 | 15 | 2,2,1,0, 2,2,1,0 | 0.67/618 |
| **8a** | $10^{-12}$ | 1 | 1 | 4 | 11 | 15 | 2 | 0.02/23 |
| **9a1** | $10^{-12}$ | 1 | 1 | 4 | 11 | 15 | 1 | 0.23/17 |
| **9a2** | $10^{-12}$ | 1 | 1 | 4 | 11 | 15 | 1 | 0.03/17 |
| **9b1** | $10^{-12}$ | 1 | 1 | 4 | 11 | 15 | 2 | 1.4/176 |
| **9b2** | $10^{-12}$ | 1 | 1 | 4 | 11 | 15 | 2 | 1.25/166 |
| **9c** | $10^{-12}$ | 1 | 1 | 4 | 11 | 15 | 1 | 0.04/1.4 |
| **9d** | $10^{-12}$ | 1 | 1 | 4 | 11 | 15 | 1 | 5.2/14 |
| **10a** | $10^{-12}$ | 1 | 1 | 4 | 12 | 15 | 1 | 1.1/26 |
| **10b** | $10^{-12}$ | 1 | 1 | 4 | 12 | 15 | 0 | 0.13/27 |
| **11a1** | $10^{-12}$ | 1 | 1 | 4 | 12 | 15 | 0 | 15.2/14.4 |
| **11a2** | $10^{-12}$ | 1 | 1 | 4 | 12 | 15 | 2 | 25.6/91.2 |
| **11b1** | $3x10^{-12}$ | 1 | 1 | 4 | 12 | 15 | 0 | 6.9/9 |
| **11b2** | $10^{-12}$ | 1 | 1 | 4 | 12 | 15 | 1 | 1.5/9 |
| **11b3** | $10^{-13}$ | 1 | 1 | 4 | 11 | 15 | 1 | 50/80 |
| **11c** | $10^{-12}$ | 1 | 1 | 4 | 12 | 15 | 3 | 3.8/29.3 |

# REFERENCES

1. Ganapol, B.D., Picca, P., Previti, A., Mostacci, D., The solution of the point kinetics equations via convergence acceleration Taylor series (CATS), ANS Topical Mtg., Knoxville, Tennessee, 2012.

2. Ganapol, B.D., A highly accurate algorithm for the solution of the point kinetics equations, *Ann. Nucl. Ener,* **62** (2013), pp. 564- 571.

3. Quintero-Leyva, B., CORE: A numerical algorithm to solve the point kinetics equations, *Ann. Nucl. Ener.*, **35** (2008), pp. 2136- 2138.

4. Sanchez, J., On the numerical solution of point reactor kinetics equations by the generalized Runge- Kutta methods, *Nucl. Sci. & Eng.*, **103** (1989), pp. 94- 99.

5. Yang, X., Jevremovic, T., Revisiting the Rosenbrock numerical solutions of the reactor point kinetics equation with numerous examples, *Nuclear Technology & Radiation Protection*, **24** (2009), 1, pp. 3- 12.

6. Li, H., Chen, W., Luo, L., Zhu, Q., A new integral method for solving the point reactor neutron kinetics equations, *Ann. Nucl. Ener.*, **36** (2009), pp. 427- 432.

7. Chao, Y., Attard, A., A resolution of the stiffness problem of reactor kinetics, *Nucl. Sci. & Eng.*, **90** (1985), pp. 40- 46.

8. Aboanber, A. E., Nahla, A. A., On Padé approximations to the exponential function and application to point kinetics equations, *Prog. Nucl. Ener.*, **44** (2004), pp. 347- 368.

9. da Nobrega, J. A. W. , A new solution of the point kinetics equations, *Nucl. Sci. & Eng.*, **46** (1971), pp. 366- 375.

10. Petersen, C. Z., Dulla, S., Vilhena, M. T. M .B., Ravetto, P., An analytical solution of the point kinetics equations with time-variable reactivity by the decomposition method, *Prog. Nucl. Ener.*, **53** (2011), pp. 1091- 1094.

11. Keepin, R.G., *Physics of Nuclear Kinetics*, Addison-Wesley, USA, 1965.

12. Kinard, M., Allen, E., Efficient numerical solution of point kinetics equations in nuclear dynamics, *Ann Nucl. Ener.*, **31** (2004), pp. 1039- 1051.

13.] P. Picca, R. Furfaro, B.D. Ganapol, A highly accurate technique for the solution of the non-linear point kinetics equations, *Ann. Nucl. Ener*, **58** (2013), pp. 43- 53.

14. Ganapol, B. D., A refined way of solving reactor point kinetics equations for imposed reactivity insertion, *Nuclear Technology & Radiation Protection*, **24** (2009), pp. 157- 165.

15. Nahla, A. A., An efficient technique for the point reactor kinetic equations with Newtonian temperature feedback effects, *Ann. Nuc. Ener.*, **38** (2011), pp. 2810- 2817.

16. Izumi, M., Noda, T., An implicit method for solving the lumped parameter reactor-kinetics equations by repeated extrapolation, *Nucl. Sci. & Eng.*, **41** (1971), pp. 299- 303.

17. Li, H., Chen, W., Luo, L., Zhu, Q., A new integral method for solving the point reactor neutron kinetics equations, *Ann. Nucl. Ener.*, **36** (2009), pp. 427- 432.

18. Hetrick, D. L., *Dynamics of Nuclear Reactors*, The University of Chicago Press, Chicago, IL, USA, 1971.

19. Kaganove, J. J., Numerical solution of the one-group space- independent reactor kinetics equations for neutron density given excess reactivity, ANL- 6132, Argonne National Laboratory, 1960.

20. McMahon, D., Pierson, A., A Taylor series solution of the reactor point kinetics equations, 2010, arXiv:1001.4100v2.

21. Nahla, A. A., Taylor series method for solving the nonlinear point kinetics equations, *Nuclear Engineering and Design*, **241**(5) (2011),1592- 1595.

22. Basken, J., Lewins, J. D., Power series solutions of the reactor kinetics equations, *Nucl. Sci. & Eng.*, **122** (1996), pp. 407- 416.

23. Vigil, J., Solution of the reactor kinetics equations by analytic continuation, *Nucl. Sci. & Eng.*, **29** (1967), pp. 392- 401.

24. Aboanber, A. E., Hamada, Y. M., PWS: An efficient code system for solving space-independent nuclear reactor dynamics, *Ann. Nucl. Ener.*, **29** (2002), pp. 2159-2172.

25. Aboanber, A. E., Hamada, Y. M., Power series solution (PWS) of nuclear reactor dynamics with Newtonian temperature feedback, *Ann. Nucl. Ener.*, **30** (2003), pp. 1111- 1122.

26. Abramowitz, M., Stegun, I.., eds., *Handbook of Mathematical Functions with Formulas, Graphs, and Mathematical Tables*, New York: Dover Publications, Tenth Printing, ISBN 978-0- 486-61272-0, (1972) pp. 16, 806, 886.

27. Sidi, A., *Practical Extrapolations Methods*, Cambridge University Press, Cambridge, (2003).

28. Hetrick, D.L., *Dynamics of Nuclear Reactors*, American Nuclear Society, 1993.

29. Schiassi,E., De Florio, M., Picca, P., Ganapol, B., Furfaro, R., Physics-informed neural networks for the point kinetics equations for nuclear reactor dynamics, *Ann. Nucl. Ener.*,**167** (2022), pp. 1- 14.

30. *https://en.wikipedia.org/wiki/General_Leibniz_rule*

31. *https://functions.wolfram.com/Bessel-TypeFunctions/BesselJ/20/ShowAll.html*.